%% file: main_final.tex
 \documentclass[final,5p,times,twocolumn,authoryear]{elsarticle}

\usepackage[T1]{fontenc}
\usepackage{lmodern}
\usepackage[utf8]{inputenc} 
\usepackage{amsmath,amssymb,mathtools}
\usepackage{graphicx}
\usepackage{booktabs}
\usepackage{longtable, multicol, lscape}
\usepackage{xcolor} 
\usepackage{enumitem}
\usepackage{orcidlink}
\usepackage{natbib}
\usepackage{url}
\usepackage{rotating}
\usepackage{multirow}

\journal{New Astronomy}

\begin{document}

\begin{frontmatter}



\title{Exploring the Near Galactic Centre: A Comprehensive Study of Bulge OCs HSC 25, HSC 37, HSC 2878 Utilising Gaia DR3 Data}



\author[1]{D. C. \c{C}{\i}nar\orcidlink{0000-0001-7940-3731}}
\ead{denizcennetcinar@gmail.com}

\author[2]{W. H. Elsanhoury\corref{cor1}\orcidlink{0000-0002-2298-4026}}
\ead{elsanhoury@nbu.edu.sa}

\author[3,4]{A. A. Haroon\orcidlink{0000-0002-8194-5836}}

\affiliation[1]{
    organization={Istanbul University, Institute of Graduate Studies in Science, Programme of Astronomy and Space Sciences},
    addressline={Beyaz{\i}t}, 
    city={Istanbul},
    postcode={34116}, 
    country={Turkey}
}

\affiliation[2]{
    organization={Physics Department, College of Science, Northern Border University},
    city={Arar},
    country={Saudi Arabia}
}

\affiliation[3]{
    organization={Astronomy and Space Science Department, Faculty of Science, King Abdulaziz University},
    city={Jeddah},
    country={Saudi Arabia}
}

\affiliation[4]{
    organization={Astronomy Department, National Research Institute of Astronomy and Geophysics (NRIAG)},
    addressline={Helwan},
    postcode={11421},
    city={Cairo},
    country={Egypt}
}

\cortext[cor1]{Corresponding author: W. H. Elsanhoury}

\begin{abstract}
We present a comprehensive photometric and kinematic study of the open clusters HSC 25, HSC 37, and HSC 2878, located in the innermost regions of the Galactic disc. Utilizing data from $Gaia$ DR3 and the \texttt{UPMASK} membership algorithm, we identify 44, 55, and 112 most probable members for HSC 25, HSC 37, and HSC 2878, respectively. The mean proper-motion components are obtained as ($-5.901 \pm 0.41$, $-6.213 \pm 0.40$), ($-3.231 \pm 0.56$, $-4.564 \pm 0.47$), and ($-3.830 \pm 0.51$, $-5.198 \pm 0.44$) mas yr$^{-1}$ for HSC 25, HSC 37, and HSC 2878, respectively. The open clusters span a broad range of evolutionary stages, with estimated ages of $\log (t/{\rm yr}) = 8.38 \pm 0.08$, $7.04 \pm 0.09$, and $9.04 \pm 0.09$, and corresponding heliocentric distances of $7.36 \pm 0.37$, $6.79 \pm 0.18$, and $6.17 \pm 0.22$ kpc. The obtained metallicities are $0.0388 \pm 0.0039$, $0.0259 \pm 0.0028$, and $0.0209 \pm 0.0023$, respectively. Total mass estimates are 135, 755, and 204 $M_{\odot}$, respectively, highlighting notable differences in stellar content across the clusters. An analysis of dynamical relaxation times suggests that HSC 25 and HSC 2878 are dynamically evolved, whereas the much younger HSC 37 is still in an early phase of dynamical evolution. The high space velocities and orbital parameters of these clusters reveal significant deviations from typical disc kinematics. HSC 25 and HSC 37 exhibit eccentric orbits and small perigalactic distances, consistent with dynamically heated or accreted origins within the Galactic bulge. In contrast, HSC 2878’s relaxed, planar orbit suggests in situ bulge membership despite its age. These findings point toward a heterogeneous dynamical origin for the clusters, with implications for star formation and evolution in the inner Milky Way.

\end{abstract}




\end{frontmatter}




\section{Introduction}\label{sec1}

Open clusters (OCs) are gravitationally bound stellar groups that emerge from the collapse of giant molecular clouds, representing fundamental building blocks of star formation in the Milky Way (MW) \citep{Lada2003, Kroupa2001}. Sharing a common origin, member stars exhibit similar chemical compositions, distances, and formation timescales, rendering OCs critical for tracing Galactic stellar populations and structure. While OCs are found across the Galactic disc, their presence in the inner regions, closer to the Galactic centre, is of particular interest due to the complex dynamical environment and the role these OCs play in probing the evolution of the central Galaxy.

The stellar populations within OCs span a broad mass range, resulting in significant diversity in luminosities, effective temperatures, and spectral types \citep{Bastian2010}. Over time, internal dynamical effects such as mass segregation, coupled with external perturbations including tidal stripping induced by the Galactic gravitational field, reshape the structure and membership of OCs \citep{BinneyTremaine2008}. Besides mapping the Galaxy’s formation history, OCs, especially those located near the Galactic centre, serve as essential calibrators for stellar evolutionary models across various age ranges, from young embedded OCs to older, dynamically evolved systems like M67 or NGC 188 \citep{Perryman1998, Cinar2024}. These characteristics firmly establish OCs as vital laboratories for studying both stellar and Galactic evolution.

The region near the Galactic centre has been investigated in the literature using a variety of Galactocentric distance thresholds \citep{Bobylev2017, Carvajal2022, Braga2019, Zoccali2024, Silva2024, Nepal2025}. In this study, we adopt a more conservative limit of 2 kpc to define the inner Galaxy. This region is known to be dynamically complex and is dominated by the influence of the central Galactic bar. Observational evidence—beginning with HI studies from the Green Bank telescope in 1978 \citep{Liszt1980} and followed by dynamical models and confirmations \citep{Binney1991}—has demonstrated that the gas and stellar kinematics within this radius are significantly shaped by a bar-like potential. Near-infrared observations \citealp[e.g.]{Blitz1991, Dwek1995} further revealed a triaxial bulge with a marked left–right asymmetry in flux distribution, emphasising the morphological and dynamical distinctiveness of this region compared to the outer disc. The OCs selected for investigation in this study, HSC 25, HSC 37, and HSC 2878, are among the nearest ($R_{\rm GC}$ < 2 kpc) known OCs to the Galactic centre \citep{Zoccali2024}. Despite their intriguing locations within the inner Galaxy, these OCs have remained largely unexamined in the literature, with no detailed analyses reported thus far. Their existence has been documented only in the \citet{Hunt2024} catalogue, which provides limited information. This lack of comprehensive study underscores a significant gap in our knowledge of stellar populations and dynamical conditions in the central regions of the MW.

The dense distribution of interstellar dust toward the Galactic centre causes intense and irregular extinction, posing a major challenge for the accurate photometric analysis of clusters in these regions. This leads to significant reddening effects, which can alter the observed photometric properties of stars and complicate the determination of fundamental cluster parameters such as distance, age, and metallicity \citep{Banks2020}. In particular, differential reddening across the field can distort colour–magnitude diagrams (CMDs) and introduce systematic uncertainties in isochrone fitting \citep{Bilir2010, Bilir2016}. Accurate modelling of these extinction effects is therefore essential to reliably characterise clusters projected onto the inner Galactic regions.

The fundamental parameters of HSC\,25, HSC\,37, and HSC\,2878, as reported in the \citet{Hunt2024} catalogue, are summarised below. HSC\,25 is located at ${\alpha_{\rm ICRS}} = 265^\circ$.392 and ${\delta_{\rm  ICRS}} = -25^\circ$.702, with proper-motion components ($\mu_{\alpha}\cos\delta$, $\mu_\delta$) of ($-5.898$, $-6.197$)~mas~yr$^{-1}$, and a trigonometric parallax of $0.053$~mas. Its estimated distance is approximately $9.89$~kpc, with a mean logarithmic age of $\log{t} = 8.903$. The cluster exhibits a mean colour excess of $E({G_{BP}-G_{RP}}) = 2.568$~mag and a $V$-band extinction of $A_{\rm V} = 5.718$~mag. 

HSC\,37 is positioned at ${\alpha_{\rm ICRS}} = 264^\circ$.355 and ${\delta_{\rm ICRS}} = -24^\circ$.406, with proper-motion components ($\mu_{\alpha}\cos\delta$, $\mu_\delta$) of ($-3.173$, $-4.548$)~mas~yr$^{-1}$, and a trigonometric parallax of $0.097$~mas. The cluster is located at a distance of approximately $6.75$~kpc and has a mean logarithmic age of $\log{t} = 7.260$. Its mean colour excess and $V$-band extinction are measured as $E({G_{BP}-G_{RP}}) = 1.944$~mag and $A_{\rm V} = 4.329$~mag, respectively.

HSC\,2878 is found at ${\alpha_{\rm ICRS}} = 256^\circ$.102 and ${\delta_{\rm ICRS}} = -37^\circ$.011, with proper-motion components ($\mu_{\alpha}\cos\delta$, $\mu_\delta$) of ($-3.837$, $-5.103$)~mas~yr$^{-1}$, and a trigonometric parallax of $0.094$~mas. Its estimated distance is approximately $7.22$~kpc, and the mean logarithmic age is determined as $\log{(t/yr)} = 9.052$. The colour excess and $V$-band extinction values for this OC are $E({G_{BP}-G_{RP}}) = 2.186$~mag and $A_{\rm V} = 4.868$~mag, respectively. 

The structure of this paper is organised as follows: Section \ref{sec2} introduces with $Gaia$ DR3 as main data sources. Section \ref{sec3} describes the analysis and results of astrometric and photometric datasets, luminosity and mass functions, dynamical, velocity elliposoid parameters, and the orbital parameters for the analysis of the OCs HSC 25, HSC 37, and HSC 2878. Section \ref{sec4} summarises the main results and highlights their implications for the understanding of stellar systems in the inner Galaxy. 

\section{Gaia DR3 Data}\label{sec2}

The present analysis relies predominantly on the third data release of the $Gaia$ mission, known as $Gaia$ DR3 \citep{Gaia2023}, as the principal resource for studying the stellar populations of three OCs: HSC 25, HSC 37, and HSC 2878. Publicly released on 13 June 2022, $Gaia$ DR3 incorporates a significant expansion in both astrometric and photometric data content compared to previous releases. It provides precise multi-band photometry in the $G$, $G_{\rm BP}$, and $G_{\rm RP}$ passbands, along with radial velocity measurements for over 33 million stars. In particular, its parallax measurements reach uncertainties as low as 0.02 to 0.03 mas for stars with $G < 15$ mag, increasing to approximately 1.3 mas at $G = 21$ mag. Proper motion uncertainties follow a similar pattern, ranging from 0.02–0.03 mas yr$^{-1}$ for brighter sources to about 1.4 mas yr$^{-1}$ for the faintest objects.

\begin{table}[h]
\caption{Mean internal photometric uncertainties ($\sigma_G$, $\sigma_{G_{\rm BP}-G_{\rm RP}}$) and astrometric uncertainties ($\sigma_{\mu}$, $\sigma_\varpi$) as a function of $G$-band magnitude for HSC~25, HSC~37, and HSC~2878. These statistics, as well as the reported star counts, are computed for all sources within the analyzed field of view, prior to any membership selection.}
\footnotesize
\begin{tabular}{cccccc}
\toprule
$G$   & $N$    & $\sigma_G$ & $\sigma_{G_{\rm BP}-G_{\rm RP}}$ & $\sigma_{\mu}$   & $\sigma_\varpi$ \\
(mag)     &        & (mag)      & (mag)                            & (mas\,yr$^{-1}$) & (mas)           \\ \hline
\multicolumn{6}{c}{HSC\,25} \\ \hline
(06,14{]} & 302    & 0.003 & 0.007    & 0.048 & 0.040  \\ 
(14,15{]} & 733    & 0.004 & 0.015    & 0.061 & 0.051  \\
(15,16{]} & 3266   & 0.004 & 0.021    & 0.089 & 0.074  \\
(16,17{]} & 8636   & 0.003 & 0.023    & 0.115 & 0.096  \\
(17,18{]} & 23,275  & 0.003 & 0.038    & 0.192 & 0.161  \\
(18,19{]} & 57,690  & 0.004 & 0.077    & 0.322 & 0.265  \\
(19,20{]} & 70,717  & 0.005 & 0.162    & 0.762 & 0.638  \\
(20,21{]} & 152,106 & 0.008 & 0.284    & 1.589 & 1.267  \\
(21,22{]} & 14,396  & 0.022 & 0.313    & 3.351 & 2.527  \\ \midrule
\multicolumn{6}{c}{HSC\,37} \\ \hline
(06,14{]} & 271    & 0.003 & 0.009    & 0.044 & 0.036  \\
(14,15{]} & 872    & 0.003 & 0.013    & 0.061 & 0.050  \\
(15,16{]} & 3062   & 0.003 & 0.014    & 0.078 & 0.064  \\
(16,17{]} & 7353   & 0.003 & 0.016    & 0.110 & 0.091  \\
(17,18{]} & 23,103  & 0.003 & 0.030    & 0.183 & 0.151  \\
(18,19{]} & 33,146  & 0.004 & 0.064    & 0.335 & 0.270  \\
(19,20{]} & 64,345  & 0.005 & 0.140    & 0.920 & 0.762  \\
(20,21{]} & 180,897 & 0.009 & 0.251    & 1.808 & 1.350  \\
(21,22{]} & 12,062  & 0.032 & 0.182    & 3.034 & 2.052  \\  \midrule
\multicolumn{6}{c}{HSC\,2878} \\ \hline
(06,14{]} & 552    & 0.003 & 0.008    & 0.043 & 0.030  \\
(14,15{]} & 960    & 0.003 & 0.009    & 0.052 & 0.036  \\
(15,16{]} & 2571   & 0.003 & 0.010    & 0.073 & 0.050  \\
(16,17{]} & 5356   & 0.003 & 0.013    & 0.111 & 0.076  \\
(17,18{]} & 1,2179  & 0.003 & 0.024    & 0.193 & 0.130  \\
(18,19{]} & 23,967  & 0.004 & 0.051    & 0.334 & 0.228  \\
(19,20{]} & 45,349  & 0.005 & 0.112    & 0.686 & 0.480  \\
(20,21{]} & 100,566 & 0.008 & 0.211    & 1.716 & 0.973  \\
(21,22{]} & 10,816  & 0.025 & 0.195    & 3.136 & 2.016   \\ \bottomrule
\end{tabular}
\label{Tab:Errors}
\end{table}

For each of the three OCs, a circular region of 20 arcminutes radius around the nominal cluster centre was used to extract data from $Gaia$ DR3. This dataset includes three-band photometric values $(G,~G_{\rm BP},G_{\rm RP})$, astrometric positions (right ascension $\alpha$ and declination $\delta$), proper-motion components $(\mu_\alpha \cos \delta,\mu_\delta)$, and trigonometric parallaxes ($\varpi$), each accompanied by their associated errors. All astrometric and photometric parameters, along with their associated uncertainties, were taken into account during the analysis. Based on the central coordinates, a total of 331,121 stellar sources were retrieved from the $Gaia$ DR3 database within a radius of 20 arcminutes centred on HSC 25, 325,111 stars around HSC 37, and 202,316 stars around HSC 2878. Table~\ref{Tab:Errors} presents the mean internal photometric uncertainties ($\sigma_G$, $\sigma_{(G_{\rm BP} - G_{\rm RP})}$) and astrometric uncertainties, including proper-motion component errors ($\sigma_{\mu_{\alpha*}}$, $\sigma_{\mu_{\delta}}$), total proper motion errors ($\sigma_{\mu}$), and trigonometric parallax errors ($\sigma_{\varpi}$), based on \textit{Gaia} DR3 data, calculated for all retrieved sources within the field of view of HSC\,25, HSC\,37, and HSC\,2878, in different $G$-band magnitude bins.

This study focuses on the OCs HSC 25, HSC 37, and HSC 2878, which were identified from the catalogue of candidate stellar associations recently compiled by \citep{Hunt2024}. This comprehensive catalogue provides updated positions, structural parameters, and preliminary classifications for a large sample of OCs, incorporating both astrometric and photometric data primarily drawn from the $Gaia$ DR3 archive. While many of the entries in the \citet{Hunt2024} catalogue have been explored in earlier studies targeting their photometric characteristics or dynamical states \citep[e.g.,][]{Elsanhoury2025, Haroon2025, YousefAlzahrani2025a, YousefAlzahrani2025b}, these three particular OCs have remained largely uncharacterised in the literature. In the present work, we conduct a detailed membership analysis and structural examination of HSC 25, HSC 37, and HSC 2878 using the high-precision data from $Gaia$ DR3. 

\section{Analysis and Results}\label{sec3}

In this study, we employed the Unsupervised Photometric Membership Assignment in Stellar Clusters (\texttt{UPMASK}) algorithm \citep{KroneMartins2014} to determine the membership probabilities ($P\geq 50\%$) of stars associated with the targeted OCs. \texttt{UPMASK} is a robust and widely-utilised tool in cluster studies, designed to identify coherent stellar populations by combining astrometric and photometric information in a statistically principled manner.

The method operates through a $k$-means clustering approach, grouping stars based on their similarity in astrometric parameters, namely proper-motion components ($\mu_{\alpha} \cos \delta$, $\mu_{\delta}$) and trigonometric parallaxes ($\varpi$), while assessing the significance of these groupings through randomisation tests. For this analysis, we utilised precise astrometric measurements from the $Gaia$ DR3 catalogue, incorporating right ascension ($\alpha$), declination ($\delta$), proper-motion components, trigonometric parallaxes, and their associated uncertainties into the algorithm. 

Following the recommendations of \citet{KroneMartins2014} for poorly populated stellar systems, and considering the intrinsically sparse nature of our targets, particularly HSC~25 and HSC~37, we adopted relatively small values for the number of stars per \textit{k}-means cluster ($n$), tailored to each system. Specifically, we set $n=7$ for HSC~25, $n=10$ for HSC~37, and $n=15$ for HSC~2878. This choice allows the algorithm to remain sensitive to small-scale local overdensities while mitigating the impact of field-star contamination. A total of 100 iterations were performed to ensure the statistical robustness of the membership assignments.

Stars satisfying the conditions of membership probability $P \geq 50\%$, confinement within the cluster’s limiting radius, and adherence to the completeness limits were classified as bona fide members. Previous applications of \texttt{UPMASK} have demonstrated its effectiveness in reliably distinguished true OC members from field contaminants \citep[e.g.,][]{CantatGaudin2020, Yontan2015, Yontan2022, Yontan2023, YontanCanbay2023, TasdemirYontan2023, TasdemirCinar2025}, and it continues to represent a cornerstone method for OC membership studies.

\subsection{Structural and Astrometric Parameters}

\begin{figure*}[ht]
\centering
\includegraphics[width=0.95\linewidth]{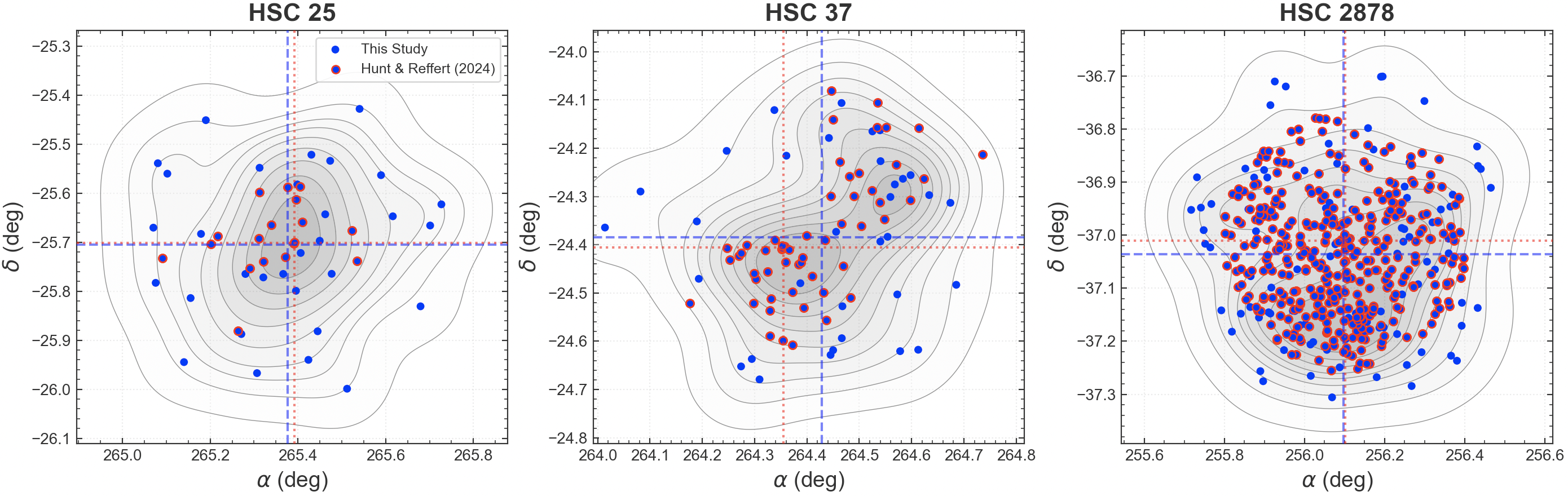}
\caption{Spatial distribution of probable members for HSC 25, HSC 37, and HSC 2878. The underlying grey contours depict the stellar density levels derived via Kernel Density Estimation (KDE) based on $Gaia$ DR3 data. Blue-filled circles represent the members identified in this study, while red open circles indicate the members adopted from \citet{Hunt2024}. The cluster centres are marked by the intersection of vertical and horizontal lines: blue dashed lines correspond to the centres determined in this work, and red dotted lines represent the coordinates given by \citet{Hunt2024}.}
\label{Fig:centers}
\end{figure*}

We applied the \texttt{UPMASK} algorithm to identify the central positions of HSC\,25, HSC\,37, and HSC\,2878 in this study \citep{CantatGaudin2020}.The cluster centres were determined using a two-dimensional Kernel Density Estimation (KDE) technique applied to the spatial distribution of the probable members. Prior to the analysis, the data were subjected to strict astrometric and photometric quality filters (e.g., $\sigma_\varpi / \varpi \leq 0.2$, proper motion error $\leq 0.5$~mas~yr$^{-1}$, and the corrected $G_{\rm BP}/G_{\rm RP}$ excess factor consistency check following \citet{Riello2021}). We employed a Gaussian kernel to estimate the probability density function of the stellar positions. To determine the optimal smoothing parameter, we adopted a bandwidth calculated via Scott's rule, which effectively highlights the density peaks while suppressing background noise. The cluster centre was identified as the location of the maximum stellar density peak. Figure~\ref{Fig:centers} displays the resulting density contours and the distribution of our members (blue filled circles). In this figure, the determined centres are marked by the intersection of vertical and horizontal blue dashed lines. For comparison, the members and central coordinates reported by \citet{Hunt2024} are indicated by red open circles and intersecting red dotted lines, respectively. The resulting coordinate values show strong agreement with the literature and are listed in Table~\ref{Tab:centers}.

\begin{table}
\caption{The revised central positions of the OCs have been derived with improved precision and are reported in both equatorial coordinates ($\alpha$, $\delta$) and Galactic coordinates ($l$, $b$).}
\centering
\footnotesize
\begin{tabular}{lcccc}
\toprule
Cluster & $\alpha$ & $\delta$ & $l$ & $b$ \\
\midrule
HSC 25 & $17^{\text{h}}\,41^{\text{m}}\,12^{\text{s}}.00$ & $-25^{\circ}\,41'\,24''$ & $2^\circ.598$ & $1^\circ.934$ \\
HSC 37 & $17^{\text{h}}\,37^{\text{m}}\,38^{\text{s}}.40$ & $-24^{\circ}\,23'\,24''$ & $3^\circ.274$ & $3^\circ.307$ \\
HSC 2878 & $17^{\text{h}}\,04^{\text{m}}\,21^{\text{s}}.60$ & $-37^{\circ}\,00'\,54''$ & $348^\circ.819$ & $2^\circ.034$ \\
\bottomrule
\end{tabular}
\label{Tab:centers}
\end{table}

\begin{figure*}[t]
\centering
\includegraphics[width=0.99\linewidth]{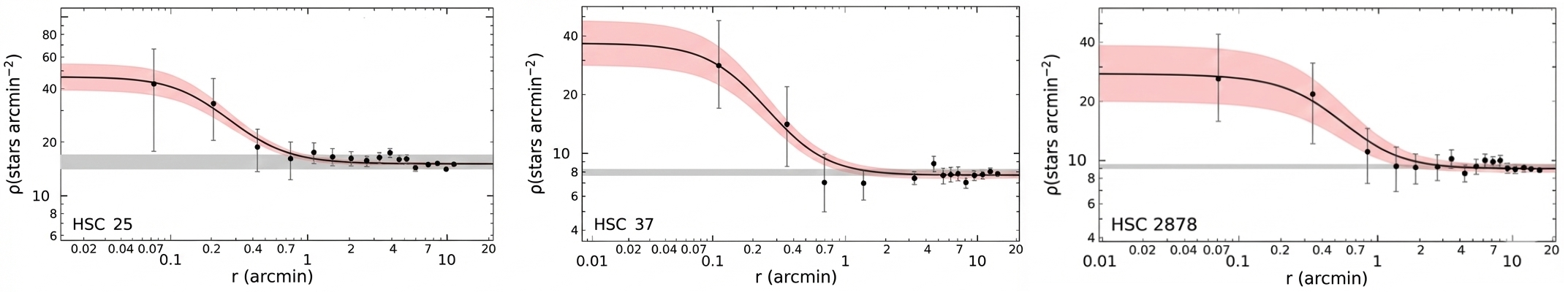}
\caption{The King profile fitting to the surface density extracted RDP parameters. The solid curve represents the best-fit empirical King profile, calculated by weighting the data points with their errors. The red-shaded band corresponds to the 1$\sigma$ confidence interval of the fit.}
\label{Fig:RDP}
\end{figure*}

To characterise the structural properties of the OCs, we constructed radial density profiles (RDPs) by measuring the stellar surface density in concentric annuli around the cluster centres. The observed RDPs were fitted with the empirical King density profile \citep{King1962}, expressed as:
\begin{equation}
\rho(r) = \rho_{\rm bg} + \rho_{0} \left[ \frac{1}{\sqrt{1 + (r/r_{\rm c})^2}} - \frac{1}{\sqrt{1 + (r_{\rm t}/r_{\rm c})^2}} \right]^2,
\end{equation}
where $\rho_{0}$ represents the central stellar density, $\rho_{\rm bg}$ is the background density, $r_{\rm c}$ is the core radius, and $r_{\rm t}$ is the tidal radius. The fitting process was performed using the weighted least-squares minimization method, where the uncertainties of individual density points were incorporated as weights ($w_i = 1/\sigma_i^2$) to provide a more realistic estimation of the parameter errors and confidence intervals.

\begin{table}
\centering
\footnotesize
\caption{The structural parameters of the investigated OCs were determined by fitting the observed surface stellar density profiles with King models. This approach enables the estimation of key parameters such as the core, limiting, and central density, offering insights into the internal structure and dynamical state of the OCs.}%
\renewcommand{\arraystretch}{1.1} 
\begin{tabular}{l|ccc}
\toprule
Parameters & HSC 25 & HSC 37 & HSC 2878 \\
\bottomrule
$r_c$ (arcmin) & 5.98$\pm$1.56 & 4.22$\pm$2.54 & 7.30$\pm$3.11 \\
$r_c$ (pc) & 22.32$\pm$5.43 & 11.80$\pm$3.96 & 19.66$\pm$5.09 \\
$r_{\rm cl}$ (arcmin) &10.09$\pm$3.66 & 7.44$\pm$3.14 & 13.54$\pm$4.23 \\
$r_{\rm cl}$ (pc) &37.64$\pm$7.06 & 20.81$\pm$5.24 & 36.47$\pm$6.95 \\
$ \rho_o$ (stars arcmin$^{-2}$) & 34.20$\pm$5.58 & 30.67$\pm$8.67 & 22.95$\pm$3.06 \\
$\rho_{bg}$ (stars arcmin$^{-2}$) & 15.08$\pm$0.82 & 7.66$\pm$0.52 & 8.92$\pm$0.26 \\
$\delta_{\rm c}$ & 3.27$\pm$0.56 & 5.01$\pm$1.46 & 3.57$\pm$0.48 \\
$C$ & 1.69$\pm$0.54 & 1.76$\pm$0.78 & 1.85$\pm$0.70 \\
\bottomrule
\end{tabular}
\label{Tab:RDP}
\end{table}

The core radii ($r_{\rm c}$) for HSC\,25, HSC\,37 and HSC\,2878 were estimated as $5.98 \pm 1.56$, $4.22 \pm 2.54$, and $7.30 \pm 3.11$ arcmin, respectively. In addition, the limiting radii ($r_{\rm cl}$), defined as the radial distance at which the cluster density merges with the background level, were determined to be $10.09 \pm 3.66$ arcmin for HSC\,25, $7.44 \pm 3.14$ arcmin for HSC\,37, and $13.54 \pm 4.23$ arcmin for HSC\,2878. The central stellar densities ($\rho_o$) were derived as $34.20 \pm 5.58$ stars arcmin$^{-2}$ for HSC\,25, $30.67 \pm 8.67$ stars arcmin$^{-2}$ for HSC\,37, and $22.95 \pm 3.06$ stars arcmin$^{-2}$ for HSC\,2878. Correspondingly, the background stellar densities ($\rho_{bg}$) were measured as $15.08 \pm 0.82$ stars arcmin$^{-2}$, $7.66 \pm 0.52$ stars arcmin$^{-2}$, and $8.92 \pm 0.26$ stars arcmin$^{-2}$, respectively.

To further assess the concentration and density enhancement of the OCs, we computed the density contrast parameter ($\delta_{\rm c}$) and the concentration parameter ($C$). The density contrast, defined by $\delta_{\rm c} = 1 + \rho_{o} / \rho_{\rm bg}$, quantifies the degree of cluster overdensity relative to the background field. The resulting values of $\delta_{\rm c}$ are $3.27 \pm 0.56$ for HSC\,25, $5.01 \pm 1.46$ for HSC\,37, and $3.57 \pm 0.48$ for HSC\,2878, suggesting significant central concentrations. The concentration parameters, calculated as $C = r_{\rm cl} / r_{\rm c}$ \citep{King1966}, were determined as $1.69 \pm 0.54$ for HSC\,25, $1.76 \pm 0.78$ for HSC\,37, and $1.85 \pm 0.70$ for HSC\,2878, indicating that the OCs possess relatively loose but distinct cores. The results of the King profile fitting applied to each open cluster are presented graphically in Figure~\ref{Fig:RDP}, while the derived structural parameters are summarised in Table~\ref{Tab:RDP}.

\begin{figure*}
\centering
\includegraphics[width=0.99\linewidth]{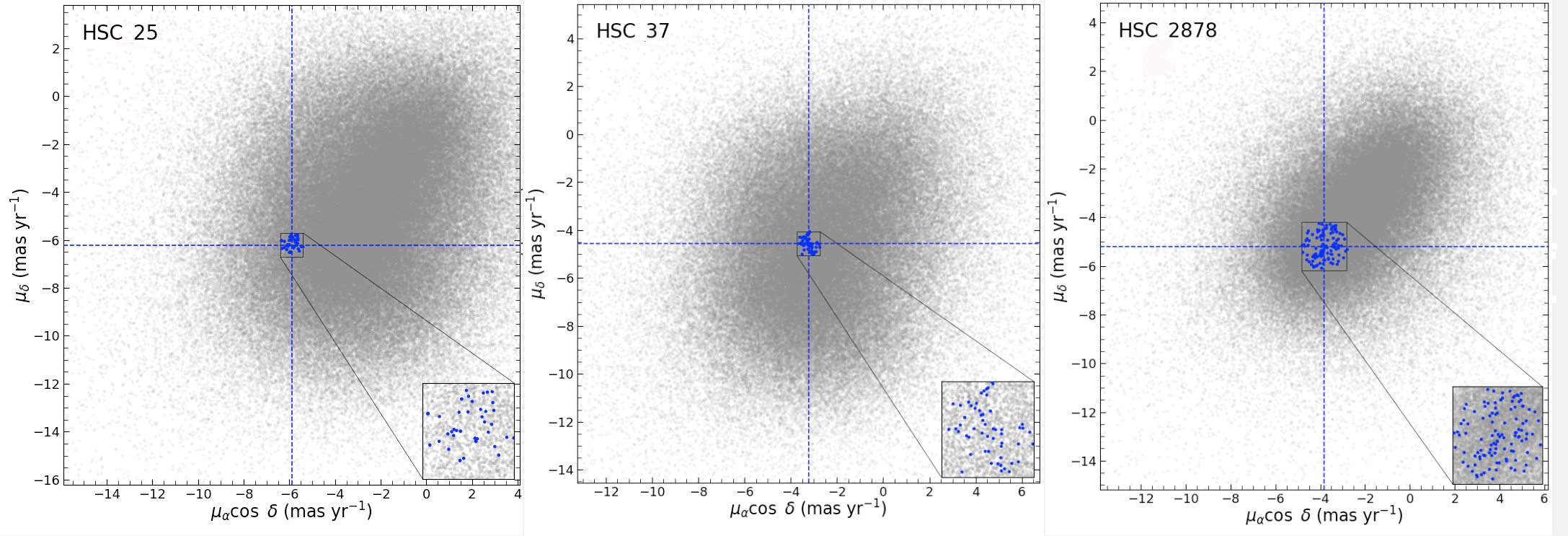}
\caption{Distributions of the mean proper-motion components ($\mu_\alpha\cos{\delta}$ and $\mu_\delta$) for the candidate OCs members.}
\label{Fig:VPD}
\end{figure*}

To investigate the kinematic properties of the clusters HSC 25, HSC 37, and HSC 2878, we constructed vector-point diagrams (VPDs) by plotting the proper-motion components $(\mu_{\alpha} \cos \delta, \mu_{\delta})$ of the member stars. As shown in the upper panels of Figure~\ref{Fig:VPD}, the mean proper-motion components of each cluster are indicated by blue dashed lines. These values were computed using stars identified as members based on their astrometric consistency. The VPDs provide a visual representation of the clusters’ kinematic distributions and support the estimation of their systemic motion. The derived mean proper-motion components for each cluster are listed in Table~\ref{Tab:maintable}.

\begin{figure}[h]
\centering
\includegraphics[width=0.9\linewidth]{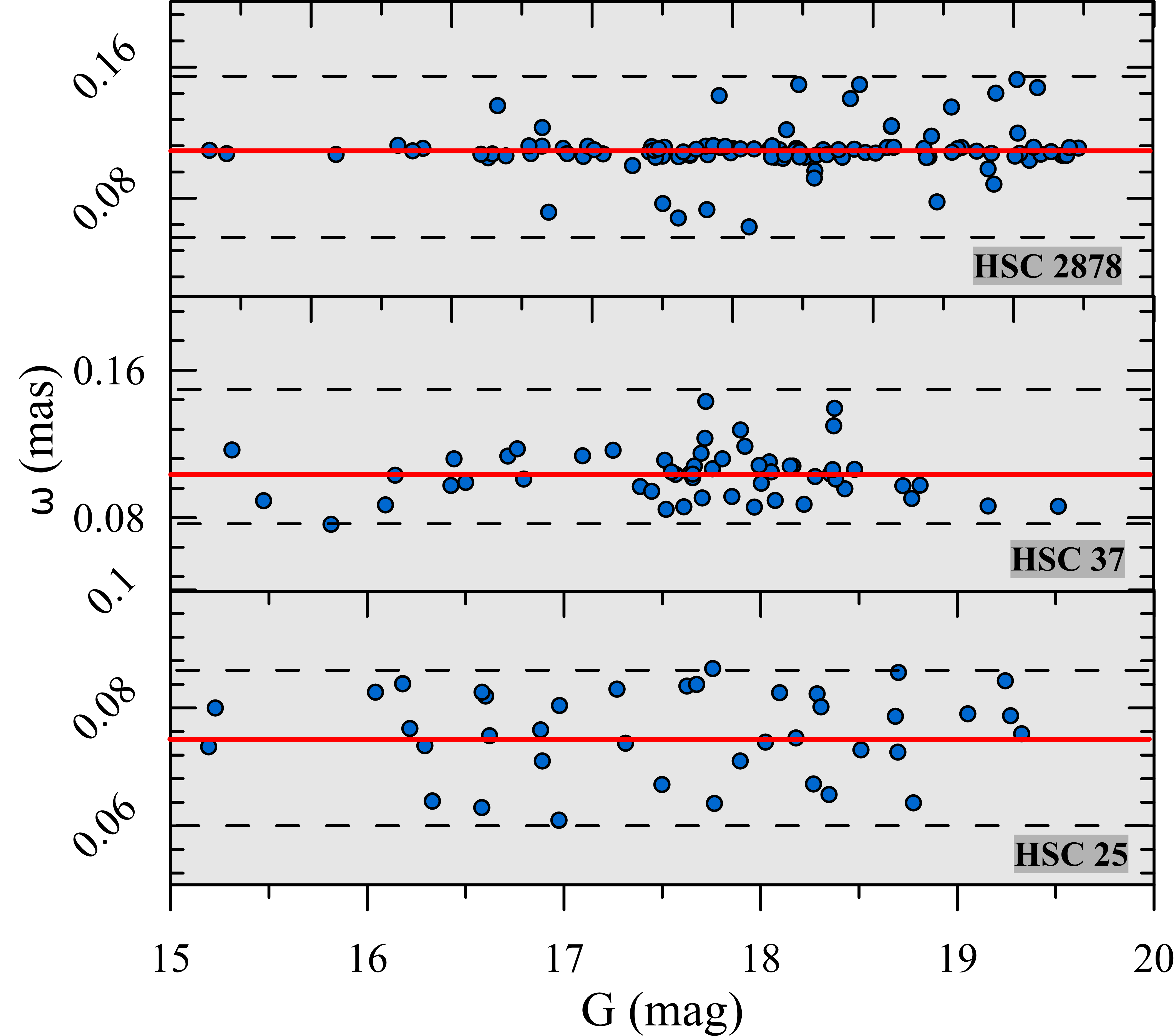}
\caption{Distribution of member stars for HSC\,25, HSC\,37, and HSC\,2878 in the trigonometric parallax ($\varpi$) versus $Gaia$ $G$-band magnitude diagram. The solid red line shows the median parallax value, and the surrounding black dashed lines show the range of values that fall within one standard deviation of the median.}
\label{Fig:Plx}
\end{figure}

The trigonometric parallaxes of the three open clusters analysed in this study were determined using $Gaia$ DR3 data by selecting member stars with relative trigonometric parallax uncertainties satisfying $\sigma_{\varpi}/\varpi \leq 0.2$. These stars were plotted on a parallax–magnitude ($\varpi$ vs. $G$) diagram, as displayed in Figure~\ref{Fig:Plx}, to assess the internal consistency of their distances. All selected stars have well-constrained trigonometric parallaxes and magnitudes brighter than $G = 19.5$. The clustering of points around a narrow trigonometric parallax range in each diagram supports their physical association.

Although the trigonometric parallax uncertainties are relatively large at the distances of the studied open clusters, the trigonometric parallaxes of the selected member stars remain within one standard deviation of the cluster mean, in line with the expected $Gaia$ DR3 measurement uncertainties. This reflects the internal consistency of the adopted member sample and the effectiveness of the membership selection in minimizing field-star contamination (via the \texttt{UPMASK} algorithm and astrometric quality cuts), rather than any improvement in the intrinsic precision of the parallax measurements.

The mean distances of the clusters were derived by inverting the individual $Gaia$ DR3 parallaxes ($\varpi_i$) of the member stars ($d_i = 1000/\varpi_i$, with $\varpi$ in milliarcseconds). To estimate the internal uncertainty of the mean distance, we computed the error of the weighted mean as follows:
\begin{equation}
\bar{d} = \frac{\sum\limits_i w_i d_i}{\sum\limits_i w_i}, \qquad \sigma_\mathrm{int} = \left( \sum\limits_i w_i \right)^{-1/2},
\end{equation}
where the weights are defined by $w_i = 1/\sigma_i^2$, and $\sigma_i$ is the uncertainty in each individual distance $d_i$. Since each distance is computed by inverting the parallax, the uncertainty is propagated accordingly using:
\begin{equation}
\sigma_i = \frac{1000 \cdot \sigma_{\varpi_i}}{\varpi_i^2}.
\end{equation}
To account for the actual spread of the distances among the cluster members, the external uncertainty was estimated using the weighted standard deviation:
\begin{equation}
\sigma_\mathrm{ext} = \sqrt{ \frac{ \sum\limits_i w_i (d_i - \bar{d})^2 }{ \sum\limits_i w_i } }.
\end{equation}

The mean trigonometric parallaxes ($\varpi$) of the cluster members were computed by taking the weighted mean of individual $Gaia$ DR3 measurements. The median values and their internal uncertainties are found to be $0.079 \pm 0.016$~mas for HSC\,25, $0.104 \pm 0.013$~mas for HSC\,37, and $0.108 \pm 0.010$~mas for HSC\,2878. These values were converted into distances using the simple inversion formula $d(pc)= 1000/\varpi$, where $\varpi$ is given in milliarcseconds (mas). Accordingly, the mean distances were estimated as $12.82 \pm2.64$~kpc for HSC\,25, $9.62 \pm 1.19$~kpc for HSC\,37, and $9.26 \pm 0.86$~kpc for HSC\,2878.

\subsection{Photometric Analysis: Reddening, Distance and Age}

In the determination of fundamental astrophysical properties of OCs, colour-magnitude diagrams (CMDs) play a pivotal role. These diagrams allow for the identification of principal evolutionary features such as the main sequence (MS), the turn-off point, and evolved stellar populations, which are critical for constraining cluster parameters. In this study, we constructed CMDs for the most probable members ($P \geq 0.5$) of HSC 25, HSC 37, and HSC 2878 using $Gaia$ photometric data. The OCs parameters, namely age, distance modulus, reddening, and metallicity, were derived by fitting {\sc PARSEC} isochrones \citep{Bressan2012}, calibrated for the $Gaia$ Early Data Release 3 (EDR3) photometric system \citep{Riello2021, Gaia2021}.

\subsubsection{Breaking the Distance--Reddening Degeneracy}

The Galactic centre region is heavily affected by interstellar dust and gas, leading to significant line-of-sight extinction and reddening. This effect becomes increasingly prominent for stellar populations located at low Galactic latitudes, where the dust column density is high due to the integrated contribution along the line of sight. OCs such as HSC~25, HSC~37, and HSC~2878, which lie in proximity to the Galactic bulge, are particularly impacted by this phenomenon. In such regions, failure to properly correct for reddening introduces substantial uncertainties in the derivation of fundamental cluster parameters such as distance, age, and metallicity. The degeneracy between distance and extinction can severely bias the results of isochrone fitting and stellar population analysis.

\begin{figure}
\centering
\includegraphics[width=0.75\linewidth]{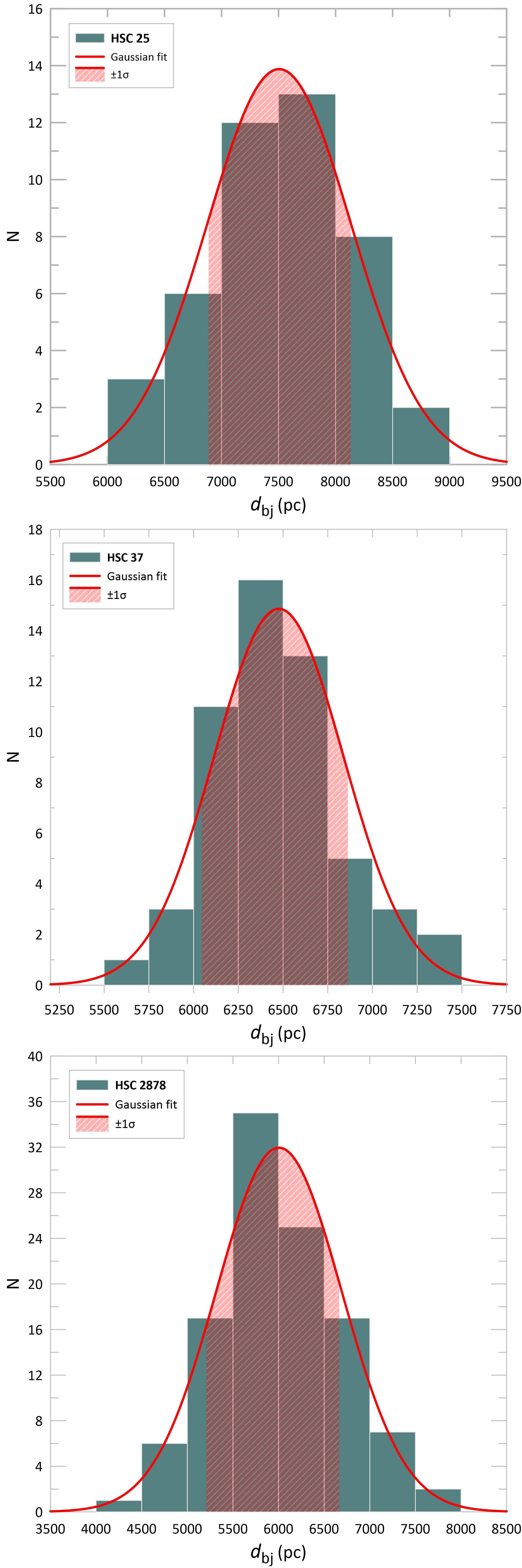}
\caption{Distance distributions of member stars in the clusters HSC~25, HSC~37, and HSC~2878 based on their individual Bailer-Jones distances ($d_{\mathrm{BJ}}$). The histograms represent the estimated distance values, while the red curve indicates the Gaussian fit applied to the distributions. The shaded region shows the $\pm1\sigma$ range around the mean, corresponding to the internal uncertainty.}
\label{Fig:Distance_BJ}
\end{figure} 

\begin{figure*}
\centering
\includegraphics[width=0.99\linewidth]{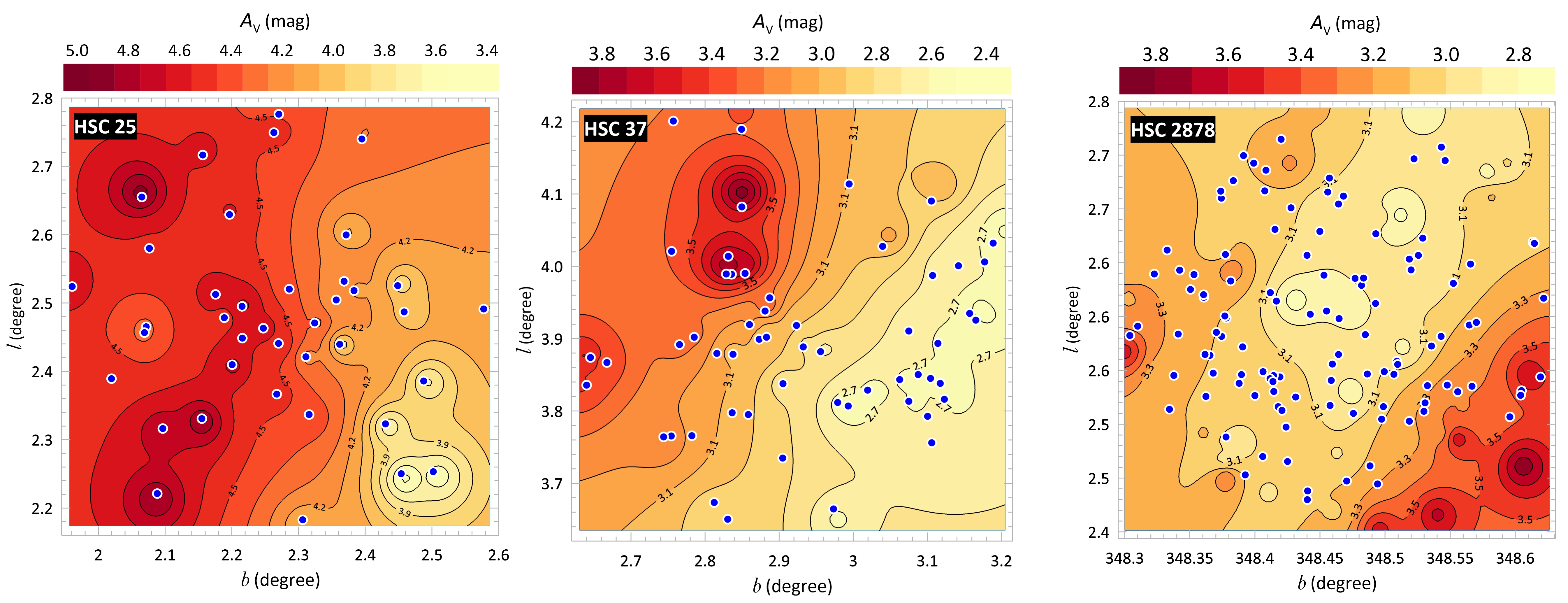}
\caption{$V$-band extinction ($A_{\rm V}$) contour maps for the fields centred on the clusters HSC 25 (left), HSC 37 (middle), and HSC 2878 (right). The extinction values are interpolated from \citet{Schlafly2011} dust maps using the IRSA interface. Blue points indicate the positions of member stars. These $A_{\rm V}$ values have been rescaled according to the individual \citet{BailerJones2021} geometric distances of the stars in each OCs field.}
\label{Fig:Reddening}
\end{figure*} 

To mitigate this degeneracy, we adopted the distance estimates provided by \citet{BailerJones2021}, who constructed a probabilistic distance catalogue for all stars in $Gaia$ EDR3 \citep{Riello2021, Gaia2021}. Their method accounts for the asymmetric and non-Gaussian nature of trigonometric parallax uncertainties, especially at small parallaxes, by applying a Bayesian inference framework with a geometric prior based on a spatial density model of the Galaxy. The catalogue offers posterior probability distributions of geometric distances for each star. In this study, we queried the distances of all confirmed member stars of OCs from this catalogue. These distances were combined for the cluster members to construct histograms of the distance distributions, as shown in Figure~\ref{Fig:Distance_BJ}. A Gaussian function was fitted to each histogram to determine the mean OC distances, while the red shaded regions in the figure indicate the corresponding $1\sigma$ uncertainties derived from the fits.

Subsequently, for each member star, we retrieved the line-of-sight total visual extinction value, $A_{\infty(V)}$, from the IRSA Galactic Dust Reddening and Extinction Service\footnote{\url{https://irsa.ipac.caltech.edu/applications/DUST/}}. This service is based on the dust maps of \citet{Schlegel1998}, with updated extinction coefficients from \citet{Schlafly2011}, and provides the integrated extinction assuming infinite distance. To derive the extinction appropriate for the actual distance of each star, we applied a vertical dust distribution model using the expression:
\begin{equation}
A_{d}(V)=A_{\infty}(V)\Biggl[1-\exp\Biggl(\frac{-\mid
d\times \sin(b)\mid}{H}\Biggr)\Biggr].
\end{equation}
where, $d$ is the heliocentric distance from \citet{BailerJones2021} and $b$ denotes the Galactic latitude. The interstellar dust distribution is modelled with a scale height ($H$) of $\pm$125 pc following \citet{Marshall2006}. We distinguished between the total dust absorption along the line of sight, $A_{\infty}(V)$, and the absorption up to the star’s distance, $A_{\rm d}(V)$, as defined in previous studies \citep{Bilir2006, Bilir2008a, Bilir2008b, Bostanci2015, Bostanci2018}. Applying this correction yielded a set of extinction values for each cluster member. As illustrated in Figure~\ref{Fig:Reddening}, each cluster exhibits noticeable differential reddening. As expected, the reddening values tend to increase toward the Galactic centre direction, reflecting the higher concentration of interstellar dust along these sightlines.

To convert visual extinction into $Gaia$ photometric system quantities, we adopted the total-to-selective extinction law of \citet{Cardelli1989}, which provides band-dependent extinction coefficients based on a standard reddening law with $R_{\rm V} = 3.1$. Specifically, the following selective extinction ratios were used, consistent with the $Gaia$ DR3 photometric response:  $A_{\rm G} / A_{\rm V} = 0.83627, A_{G_{\rm BP}} / A_{\rm V} = 1.08337$, and  $A_{G_{\rm RP}}/A_{\rm V} = 0.63439$ \citealp[see also,]{Canbay2023}. The $Gaia$ colour excess $E(G_\text{BP} - G_\text{RP})$ was then estimated using the empirical relation:
\begin{equation}
E(G_\text{BP} - G_\text{RP}) = \frac{A_{\rm G}}{1.8626},
\end{equation}
where $A_{\rm G}$ denotes the extinction in the $Gaia$ $G$-band, computed from the relation $A_{\rm G} = 0.83627 \times A_{\rm V}$. This transformation enabled the determination of individual reddening values in $Gaia$ colours for each cluster member, which were subsequently incorporated into the isochrone fitting procedure. As a result of this procedure, $A_{\rm V}$ values for HSC 25, HSC 37, and HSC 2878 were estimated to be $4.43 \pm 0.29$, $2.93 \pm 0.29$, and $3.41 \pm 0.41$ mag, respectively. Using the adopted extinction transformations, $E(G_\text{BP} - G_\text{RP})$ were calculated as $1.99 \pm 0.13$, $1.32 \pm 0.13$, and $1.54 \pm 0.18$ mag, respectively. Taking into account that each cluster member may be affected by differential reddening, we adopted individual reddening values for each star and incorporated them accordingly in the isochrone fitting analysis.

The CMDs in the $G \times (G_{\rm BP} - G_{\rm RP})$ plane were used to perform the isochrone fitting for each cluster, focusing on the most reliable stellar members. A visual fitting method was adopted, emphasising the alignment of the isochrones with the observed MS and evolved stars \citep{Bisht2022a, Bisht2022b, Bisht2025}. Prior to fitting, the metallicity [Fe/H] of each cluster was converted into the mass fraction $z$ by employing the analytic relations provided by Bovy\footnote{https://github.com/jobovy/isodist/blob/master/isodist/Isochrone.py} for {\sc PARSEC} models:
\begin{equation} z_{\rm x} = 10^{\rm [Fe/H]} \times \left(\frac{z_{\odot}}{1-0.248-2.78 \times z_{\odot}}\right), \end{equation} 
\begin{equation} z = \frac{z_{\rm x} - 0.2485 \times z_{\rm x}}{2.78 \times z_{\rm x} + 1}. \nonumber \end{equation}
where the adopted solar metallicity is $z_{\odot} = 0.0152$ \citep{Bressan2012}. For HSC 25, HSC 37, and HSC 2878, the metallicities were determined as $z = 0.0388 \pm 0.0039$, $0.0259 \pm 0.0028$, and $0.0209 \pm 0.0023$, respectively. Isochrone fitting yielded OCs ages of $\log (t/{\rm yr}) = 8.38 \pm 0.08$ for HSC 25, $\log (t/{\rm yr}) = 7.04 \pm 0.09$ for HSC 37, and $\log (t/{\rm yr}) = 9.04 \pm 0.09$ for HSC 2878.

\begin{figure*}[ht]
\centering
\includegraphics[width=0.75\linewidth]{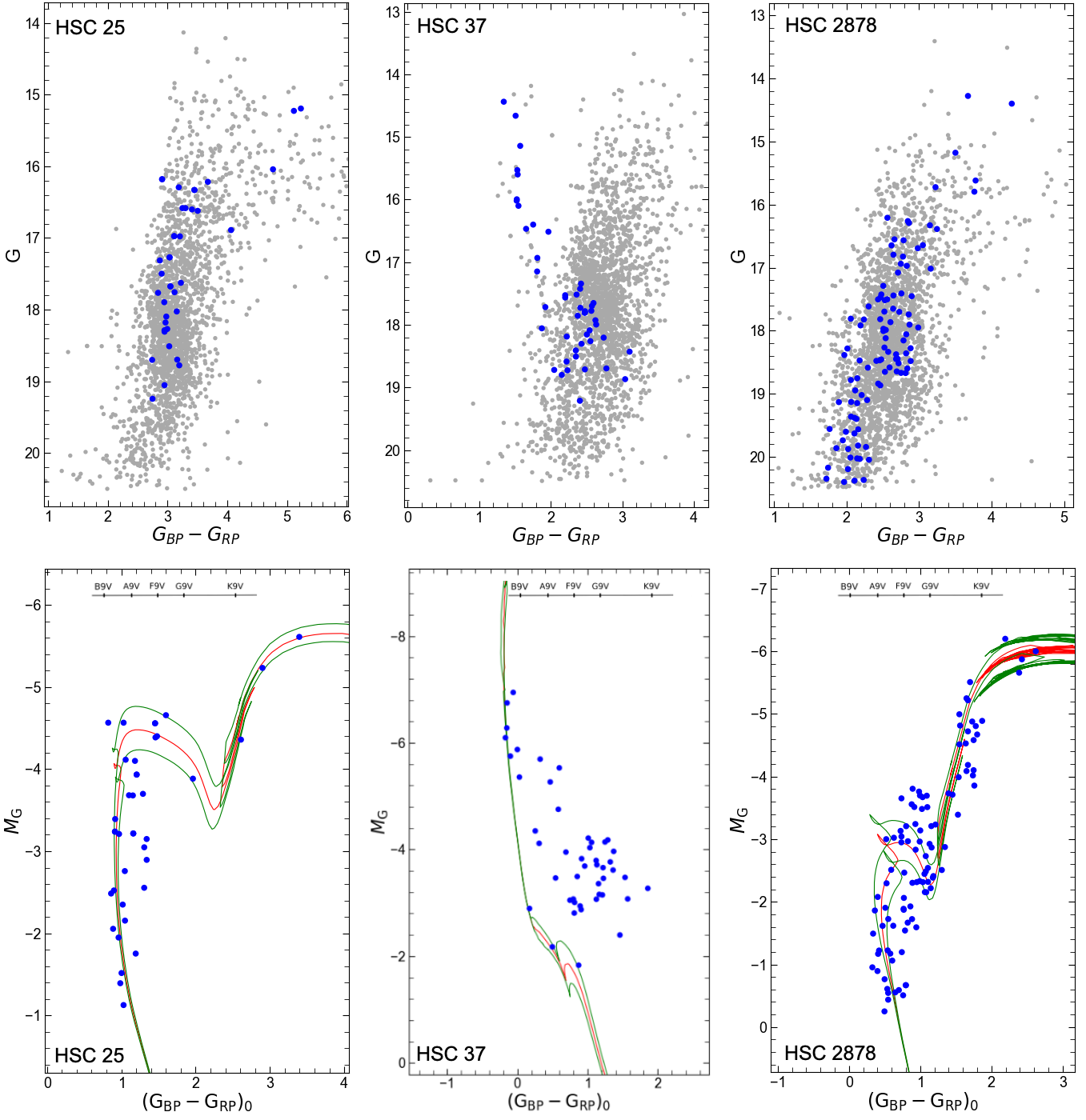}
\caption{colour--magnitude diagrams (CMDs) for the clusters HSC\,25, HSC\,37, and HSC\,2878. Top row: Observed $G$ versus $(G_{\rm BP} - G_{\rm RP})$ distributions including both member (blue) and field stars (gray), without correction for interstellar reddening. Bottom row: Reddening-corrected CMDs showing absolute magnitude $M_G$ versus intrinsic colour $(G_{\rm BP} - G_{\rm RP})_0$ for the same clusters. The red curves represent the best-fitting PARSEC isochrones, whereas the surrounding green curves delineate the uncertainty bounds associated with the fitting process.}
\label{Fig:CMDs}
\end{figure*} 

From the isochrone fitting, the distance moduli were derived as $(m-M) = 18.04 \pm 0.09$ mag for HSC 25, $16.62 \pm 0.06$ mag for HSC 37, and $16.82 \pm 0.08$ mag for HSC 2878, corresponding to distances of $7.36 \pm 0.37$, $6.79 \pm 0.18$, and $6.17 \pm 0.22$ kpc, respectively. All obtained fundamental astrometric and astrophysical parameters within HSC 25, HSC 37, and HSC 2878 are listed here with Table~\ref{Tab:maintable}. The spectral type classes \citep{Pecaut2013} shown in Figure~\ref{Fig:CMDs}, the pre-MS region is visible before the early F-type stars and indicates the stage at which these stars reach the MS for HSC 37. Although the CMD has been corrected for the line-of-sight reddening, the stars in this evolutionary phase are still embedded in a relatively dusty environment. This residual internal extinction, likely associated with the cluster’s natal material, is still discernible in the CMD as a slight colour excess among the pre-MS members.

The isochrone distances are in general agreement with those derived from the mean distances of \citet{BailerJones2021}, yielding $d_{\rm BJ} = 7.51 \pm 2.35$ kpc for HSC 25, $6.46 \pm 2.92$ kpc for HSC 37, and $5.94 \pm 2.42$ kpc for HSC 2878. However, a noticeable discrepancy is observed for HSC 25, where the trigonometric parallax-based distance exceeds the isochrone distance. This difference can plausibly be attributed to the significant interstellar reddening effects associated with the proximity of HSC 25 to the Galactic centre, which may lead to an overestimation of distances when relying solely on trigonometric parallax measurements. 

\begin{table*}[h]
\centering
\footnotesize
\caption{A detailed comparison of the astrophysical and photometric properties determined for HSC 25, HSC 37, and HSC 2878 within the scope of this investigation is presented.}
\label{Tab:maintable}
\renewcommand{\arraystretch}{1.1} 
\begin{tabular}{l|ccc}
\toprule
Parameters   & HSC 25 & HSC 37 & HSC 2878 \\ 
\midrule
No. of members & 44  & 55  & 112 \\ \hline
&\multicolumn{3}{c}{Astrometric Parameters} \\\hline
$\mu_\alpha \cos \delta$ (mas yr$^{-1}$) & -5.901 $\pm$ 0.41 & -3.231 $\pm$ 0.56 & -3.830 $\pm$ 0.51 \\
$\mu_\delta$ (mas yr$^{-1}$) & -6.213 $\pm$ 0.40 & -4.564 $\pm$ 0.47 & -5.198 $\pm$ 0.44 \\
$\boldsymbol{\varpi}$ (mas) & 0.079 $\pm$ 0.016 & 0.104  $\pm$  0.013 & 0.108  $\pm$  0.010 \\
$\boldsymbol{d_{\varpi}} (kpc) $ & 12.82 $\pm$ 2.64 & 9.62 $\pm$ 1.19 & 9.26 $\pm$ 0.86 \\ 
$\boldsymbol{d_{\rm BJ}}$ (kpc) & 7.51 $\pm$ 2.35 & 6.46 $\pm$ 2.92 & 5.94 $\pm$ 2.42 \\\hline
&\multicolumn{3}{c}{Astrophysical Parameters} \\\hline
$A_{\text{V}}$ (mag) & 4.43 $\pm$ 0.29 & 2.93 $\pm$ 0.29 & 3.41 $\pm$ 0.41 \\
$E(B-V)$ (mag) & 1.41 $\pm$ 0.09 & 0.94 $\pm$ 0.09 & 1.09 $\pm$ 0.13 \\
$E(G_\text{BP} - G_\text{RP})$ (mag) & 1.99 $\pm$ 0.13 & 1.32 $\pm$ 0.13 & 1.54 $\pm$ 0.18 \\
$d_{\rm iso}$ (kpc) & 7.36 $\pm$ 0.37 & 6.79 $\pm$ 0.18 & 6.17 $\pm$ 0.22 \\
$(m-M)$ (mag) & 18.04 $\pm$ 0.09 & 16.62 $\pm$ 0.06 & 16.82 $\pm$ 0.08 \\
Z & 0.0388 $\pm$ 0.0039 & 0.0259 $\pm$ 0.0028 & 0.0209 $\pm$ 0.0023 \\
$\log (t/{\rm yr})$   & 8.38 $\pm$ 0.08 & 7.04 $\pm$ 0.09 & 9.04 $\pm$ 0.09 \\
$X_\odot$ (kpc) & 7.33 $\pm$ 0.34 & 6.64 $\pm$ 0.18 & 6.16 $\pm$ 0.22 \\
$Y_\odot$ (kpc) & 0.37 $\pm$ 0.02 & -1.36 $\pm$ 0.04 & 0.25 $\pm$ 0.01 \\
$Z_\odot$ (kpc) & 0.51 $\pm$ 0.02 & 0.31 $\pm$ 0.01 & 0.27 $\pm$ 0.01 \\
$R_\text{gc}$ (kpc) & 0.76 $\pm$ 0.25 & 1.92 $\pm$ 0.09 & 1.86 $\pm$ 0.21 \\

\bottomrule
\end{tabular}
\end{table*}

To explore the spatial distribution of the OCs within the MW, the Galactocentric distance ($R_\text{gc}$) was computed using the following formula:
\begin{equation}
R_{\rm gc} = \sqrt{R_0^2 + d^2 \cos^2 b - 2 R_0 d \cos b \cos l}
\end{equation}
where $R_{\circ}$ is the distance from the Sun to the Galactic centre set at $R_{\rm gc}=8$ kpc \citep{Bovy2012, Bovy2015}, $d$ is the distance to the cluster, and $l$ and $b$ represent the Galactic longitude and latitude, respectively. In addition, the Galactocentric Cartesian coordinates $(X, Y, Z)_\odot$ for each cluster were calculated. In this context, the vector $X$ is directed towards the Galactic centre, $Y$ towards the direction of Galactic rotation, and $Z$ towards the Galactic north pole. The resulting positions for HSC 25, HSC 37, and HSC 2878 are listed in Table~\ref{Tab:maintable}.

Furthermore, the Galactocentric distances were calculated as $R_{\rm gc} = 0.76 \pm 0.25$ kpc for HSC 25, $1.92 \pm 0.09$ kpc for HSC 37, and $1.86 \pm 0.21$ kpc for HSC 2878. Compared to the values reported by \citet{Hunt2024}, HSC 25 and HSC 2878 are found to be closer to the Galactic centre, while HSC 37 appears slightly farther.

\subsection{Dynamical and Kinematical Structure}\label{sec6}

The dynamical and kinematical characterisation of OCs are essential for understanding their formation, evolution, and interaction with the Galactic environment. The internal dynamical state reflects processes such as mass segregation, energy equipartition, and tidal stripping, while the kinematical properties provide critical information about the OCs' origins and their trajectories within the MW. By examining relaxation times, space velocities, Solar motion components, and orbital parameters, it is possible to reconstruct the dynamical histories of OCs and infer their membership in distinct Galactic populations. Such analyses contribute not only to the detailed mapping of the Galactic structure but also to the broader understanding of star cluster evolution under the influence of external gravitational perturbations.

\subsubsection{The Dynamical Relaxation Time}

The Luminosity Function (LF) is a useful tool for examining the distribution of stellar brightness within a cluster, expressed as the number of stars per absolute magnitude interval \citep{Haroon2017}. In this study, $G$-band LFs were constructed for stars with absolute magnitudes derived from the reddening-corrected distance moduli $(m-M)_0$ determined for each OC, and are shown in Figure \ref{Fig:LF}. 

{Due to the limited number of main-sequence (MS) stars in HSC\,2878 and the insufficient MS population in HSC\,25 and HSC\,37, a reliable slope determination for the Initial Mass Function (IMF) could not be performed. Therefore, stellar masses were estimated on a star-by-star basis using the mass–luminosity relations provided by the best-fitting isochrones. For each high-probability member star, the reddening-corrected absolute magnitudes and colours were matched to the corresponding points along the isochrone, and the stellar mass was directly assigned. For evolved stars, individual masses were inferred according to their evolutionary stage on the isochrones.

Using this approach, we obtained observed stellar masses of approximately $135,M_\odot$ for HSC\,25, $755,M_\odot$ for HSC\,37, and $204,M_\odot$ for HSC\,2878, including both main-sequence and evolved members. These values represent the observed cluster masses, derived solely from the confirmed member stars, without any extrapolation toward lower, unobserved stellar masses. A summary of the derived parameters, including the observed cluster mass ($M_{\rm obs}$) and the mean stellar mass $\langle M \rangle$, is given in Table~\ref{tab:full_parameters}.

\begin{figure*}
\centering
\includegraphics[width=0.99\linewidth]{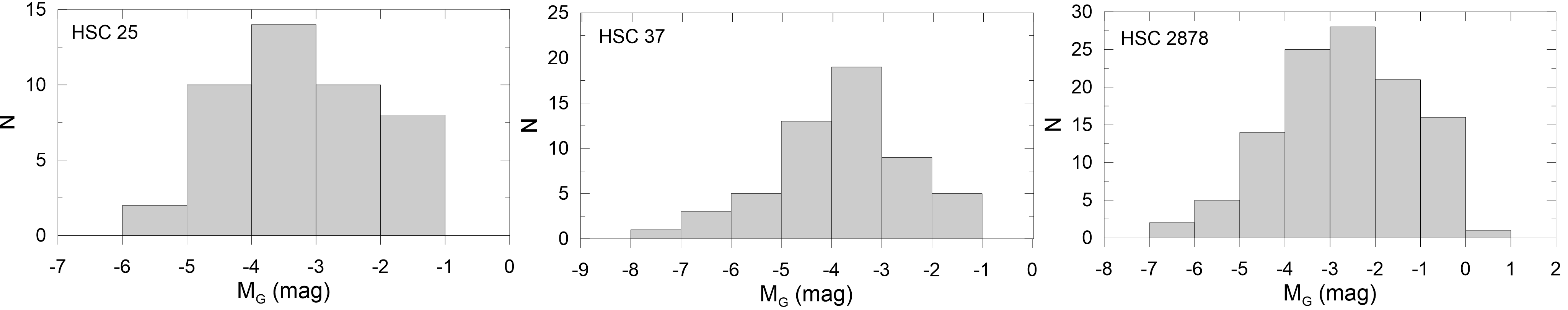}
\caption{The luminosity function frequency distributions in the absolute magnitude $G$-band ($M_G$) have been derived for the three OCs examined in this study.}
\label{Fig:LF}
\end{figure*} 

The dynamical relaxation time ($T_\text{relax}$) characterises the period required for a star cluster to reach dynamical equilibrium, where the internal stellar motions balance the competing processes of contraction and potential disruption. Estimating this timescale is essential for assessing the dynamical maturity of a cluster, revealing whether it has achieved a stable configuration or remains dynamically young \citep{Haroon2017}. A short $T_\text{relax}$ implies that the cluster has undergone significant internal dynamical evolution, whereas a long $T_\text{relax}$ suggests a less dynamically evolved system.

Following the prescription of \citet{SpitzerHart1971}, the dynamical relaxation time is calculated using the relation: 
\begin{equation} \label{Eq: 6} T_\text{relax} = \frac{8.9 \times 10^5 N^{1/2} R_\text{h}^{3/2}}{\langle M_C \rangle^{1/2} \log(0.4N)}, \end{equation} 
where $N$ is the total number of stars in the cluster, $\langle M_C \rangle$ denotes the mean stellar mass, and $R_{\rm h}$ is the half-mass radius expressed in parsecs. The half-mass radius was derived according to the transformation proposed by \citet{Sableviciute2006}: \begin{equation} R_{\rm h} = 0.547 \times r_{\rm c} \times \left( \frac{r_{\rm t}}{r_{\rm c}} \right)^{0.486}, \end{equation} 
where $r_{\rm c}$ is the core radius and $r_{\rm t}$ is adopted as the limiting radius $r_{\rm cl}$, obtained from the cluster RDP as the radius where the fitted King profile meets the background density, following the method described by \citep{Ozturk2025}. This substitution is commonly employed in cases where the formal tidal radius cannot be robustly constrained observationally. Based on these parameters, the corresponding dynamical relaxation times are estimated as 156 $\pm$ 54 Myr for HSC 25, 31 $\pm$ 15 Myr for HSC 37, and 214 $\pm$ 76 Myr for HSC 2878. It should be noted that the mean stellar mass used in the relaxation time calculation is derived solely from the observed member stars and does not include stars below the detection limit. As a result, the reported $T_{\rm relax}$ values should be regarded as upper limits, since the inclusion of lower-mass stars would reduce the mean stellar mass and lead to shorter relaxation times.

\subsubsection{Determination of Galactic Velocity Components}

The space-velocity components ($V_x$, $V_y$, $V_z$) for individual stars were derived based on their proper-motion components, radial velocities, and distances utilising the standard relations presented in \citet{Haroon2025}. Radial velocity is a key parameter for accurately reconstructing the orbital trajectories of stellar systems around the Galactic centre. In this study, the mean radial velocities of HSC 25, HSC 37, and HSC 2878 were determined individually. For each cluster, the most probable members with available radial velocity measurements in {\it Gaia} DR3 were selected, resulting in 2, 5, and 6 stars, respectively. The mean radial velocities were calculated according to the procedure described by \citet{Carrera2022}, yielding $V_{\rm R} = -101.09 \pm 35.72$ km s$^{-1}$ for HSC 25, $V_{\rm R} = 41.04 \pm 29.08$ km s$^{-1}$ for HSC 37, and $V_{\rm R} = -124.64 \pm 19.43$ km s$^{-1}$ for HSC 2878. It is worth noting that while the radial velocity of HSC 25 is not provided in the \citet{Hunt2024} catalogue, the radial velocities derived for HSC 37 and HSC 2878 in this work are in good agreement with the values reported therein.

To interpret the kinematics within the Galactic frame, the space velocity vectors were transformed from equatorial to Galactic coordinates ($U$, $V$, $W$), following the transformation matrix outlined by \citet{Haroon2025}. Here, $U$ is defined positive toward the Galactic centre $V$ in the direction of Galactic rotation, and $W$ toward the North Galactic Pole. The mean Galactic space-velocity components ($\overline{U}$, $\overline{V}$, $\overline{W}$) were subsequently obtained by averaging over the OC members. The distributions of individual ($U$, $V$, $W$) components for the OC members are presented in Figure~\ref{Fig:UVW}, while the computed mean values for each OC are summarised in Table~\ref{tab:full_parameters}.

The estimation of the solar motion components ($U_{\odot}$, $V_{\odot}$, $W_{\odot}$) relies on the Galactic space velocities of well-characterised stellar groups that can act as reference frames for the Sun’s kinematics within the MW. In this study, the mean Galactic space-velocity components ($\overline{U}$, $\overline{V}$, $\overline{W}$) obtained for HSC 25, HSC 37, and HSC 2878 were utilised to infer the solar motion parameters.

\begin{figure*}
\centering
\includegraphics[width=0.99\linewidth]{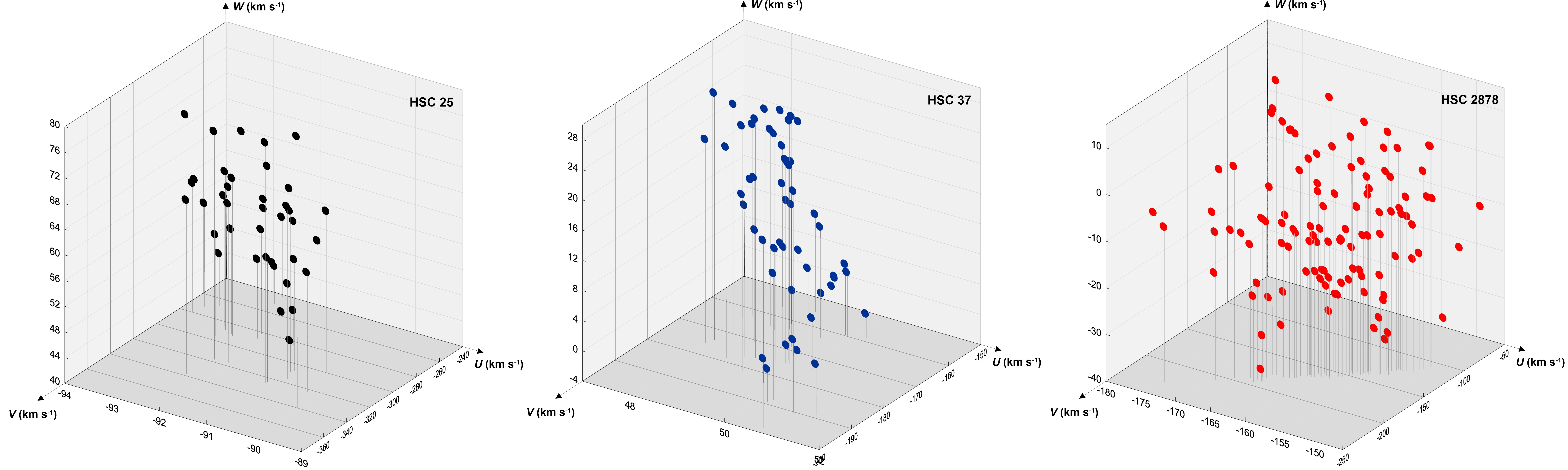}
\caption{Velocity dispersion profiles of HSC 25, HSC 37, and HSC 2878 projected onto the Galactic coordinate axes, illustrating the distribution of their spatial motion components.}
\label{Fig:UVW}
\end{figure*} 

To investigate the intrinsic space motions of the HSC\,25, HSC\,37, and HSC\,2878 within the Galactic context, we transformed their heliocentric velocity vectors into the Local Standard of Rest (LSR) frame. This transformation corrects for the peculiar motion of the Sun, for which we adopted the values determined by \citet{Coskunoglu_2011}, namely $(U, V, W)_\odot = (8.83 \pm 0.24,\ 14.19 \pm 0.34,\ 6.57 \pm 0.21)$~km\,s$^{-1}$. In this convention, $U$ is directed towards the Galactic centre, $V$ follows the direction of Galactic rotation, and $W$ is oriented towards the north Galactic pole.

To further refine the space velocity components, we applied corrections for the differential rotation of the Galaxy using the classical Oort constants $A$ and $B$. The corrections to the velocity components $U$ and $V$ are given by:
\begin{equation}
\Delta U = -A \, d \sin(2l) \cos^2 b,
\end{equation}
\begin{equation}
\Delta V = -A \, d \cos(2l) \cos^2 b - B \, d \cos b,
\end{equation}
where $d$ is the heliocentric distance, and $l$ and $b$ are the Galactic longitude and latitude of the cluster, respectively. We adopted the Oort constants values as $A = 15.3 \pm 0.4$ km\,s$^{-1}$kpc$^{-1}$ and $B = -11.9 \pm 0.4$ km\,s$^{-1}$kpc$^{-1}$ \citep{Feast1997}. These differential rotation corrections allow us to isolate the intrinsic motions of the clusters relative to the Local Standard of Rest more precisely by removing the systematic effects of Galactic rotation.

The velocity components of the OCs in the LSR frame were calculated by subtracting the solar motion vector from their observed heliocentric velocity vectors. This correction enables an accurate determination of the OCs’ motions relative to the mean motion of stars in the Solar neighbourhood. The total space velocity relative to the LSR, denoted as $S_{\rm LSR}$, is computed from the corrected velocity components using the relation:
\begin{equation}
S_{\rm LSR} = \sqrt{U_{\rm LSR}^2 + V_{\rm LSR}^2 + W_{\rm LSR}^2}.
\end{equation}

Applying this formulation, we obtained LSR-relative total space velocities of $S_{\rm LSR} = 279.23 \pm 21.28$,~km~$^{-1}$ for HSC\,25, $S_{\rm LSR} =151.64 \pm 15.26$~km~$^{-1}$ for HSC\,37, and $S_{\rm LSR} = 225.79 \pm 17.94 $~km~s$^{-1}$ for HSC\,2878. The derived $S_{\rm LSR}$ values for HSC\,25, HSC\,37, and HSC\,2878 are significantly higher than the typical velocities associated with thin-disc stellar populations. Given their positions and kinematics, these open clusters are identified as bulge clusters rather than typical thin-disc members. Their elevated space velocities, exceeding 150~km~s$^{-1}$ in all cases, reflect the dynamically hotter environment of the Galactic bulge. Such kinematic signatures imply that these OCs may have originated in a dynamically hotter environment or could be remnants of accreted systems. Their high total velocities support this interpretation and point to a more complex formation or evolutionary history compared to classical young OCs \citep{Leggett_1992}.

\subsubsection{Orbit Parameters and Galactic Population}

To analyse the orbital characteristics of HSC\,25, HSC\,37, and HSC\,2878, we utilised both axisymmetric and non-axisymmetric gravitational models implemented in the {\sc galpy}\footnote{\url{https://galpy.readthedocs.io/en/v1.5.0/}} library, a widely adopted Python toolkit for Galactic dynamics studies. As the baseline axisymmetric potential, we adopted {\sc MWPotential2014}, developed by \citet{Bovy2015}, which provides a self-consistent and observationally calibrated model of the Milky Way’s gravitational field. This potential comprises three main components: a power-law bulge with an exponential cut-off, a flattened disc following the Miyamoto–Nagai formulation, and a spherical dark matter halo modelled by the Navarro–Frenk–White (NFW) profile.

The total gravitational potential of this axisymmetric model is expressed as
\begin{equation}
\Phi_{\rm total}(R, z) = \Phi_{\rm bulge}(r) + \Phi_{\rm disc}(R,z) + \Phi_{\rm halo}(r),
\end{equation}
where $r = \sqrt{R^2 + z^2}$ denotes the Galactocentric spherical radius. {\sc MWPotential2014} is particularly well-suited for orbit integrations in the outer and intermediate regions of the Galaxy due to its ability to reproduce key observables, such as the Galactic rotation curve and local circular velocity.

\begin{figure}[ht]
\centering
\includegraphics[width=0.8\linewidth]{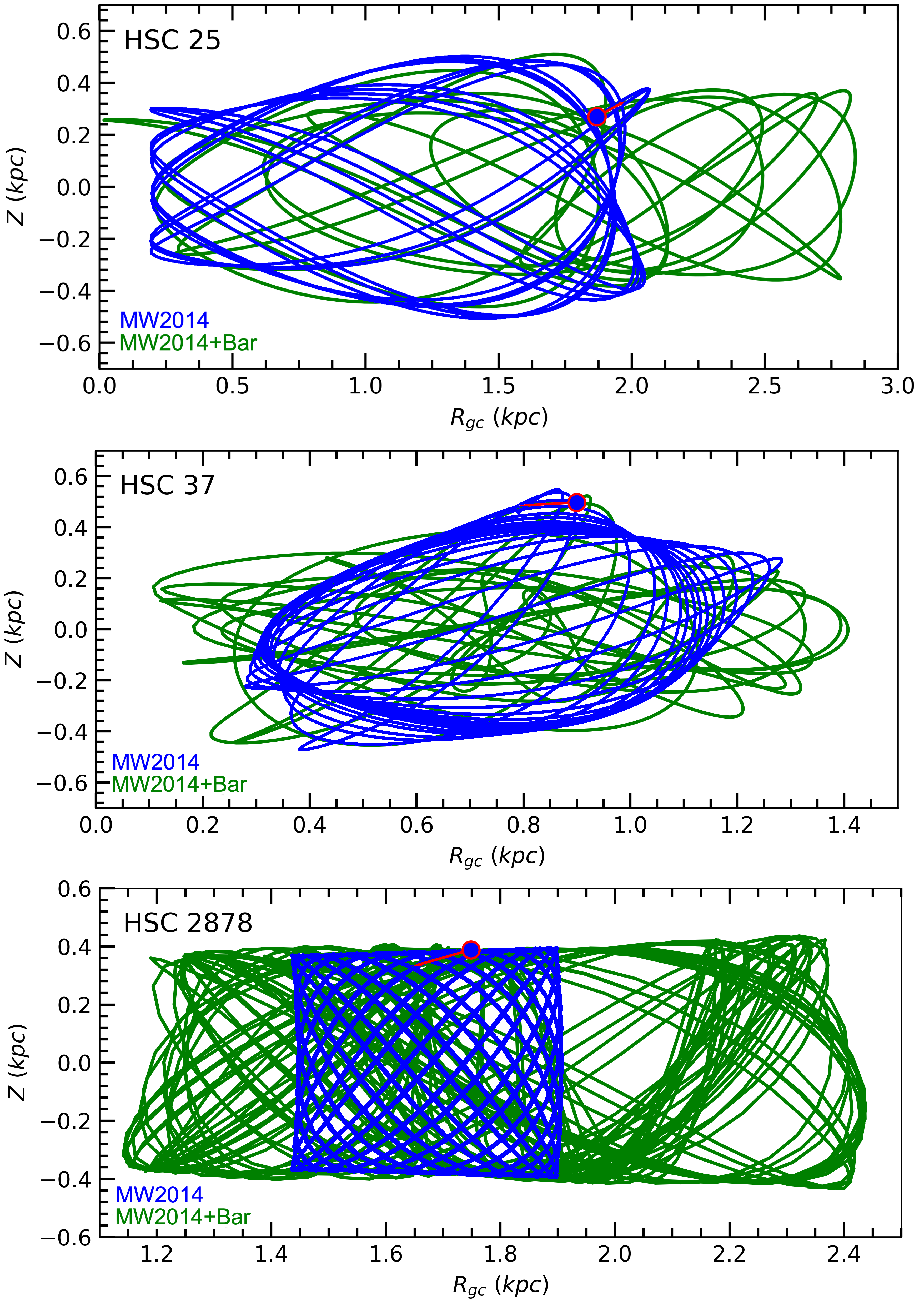} 
\caption{Galactic orbits of HSC\,25, HSC\,37, and HSC\,2878 projected in the $Z$ vs. $R_{\rm gc}$ plane. Blue lines show orbits integrated using the axisymmetric \texttt{MWPotential2014}, while green lines include the additional effect of a non-axisymmetric bar component (\texttt{DehnenBarPotential}). Arrows indicate the direction of motion. The inclusion of the Galactic bar results in noticeable deviations, particularly near perigalactic passages.}
\label{Fig:Orbits} 
\end{figure}

However, for objects located in the inner Galaxy ($R_{\rm gc} < 2$\,kpc), the axisymmetric approximation becomes insufficient due to the prominent influence of the Galactic bar. To account for this, we further included the non-axisymmetric {\tt DehnenBarPotential}, which models the bar as a rotating quadrupole perturbation with time-dependent amplitude. Its potential takes the general form:
\begin{equation}
\Phi_{\rm bar}(R, \phi, t) = A(t) \cos[2(\phi - \Omega_{\rm bar} t)] \times f(R),
\end{equation}
where $\Omega_{\rm bar}$ is the bar's pattern speed, and $A(t)$ describes the secular growth of the bar. The radial function $f(R)$ defines how the strength of the perturbation varies across the disc and decays beyond a characteristic bar radius $R_{\rm bar}$.

\begin{figure*}[ht]
\centering
\includegraphics[width=0.8\linewidth]{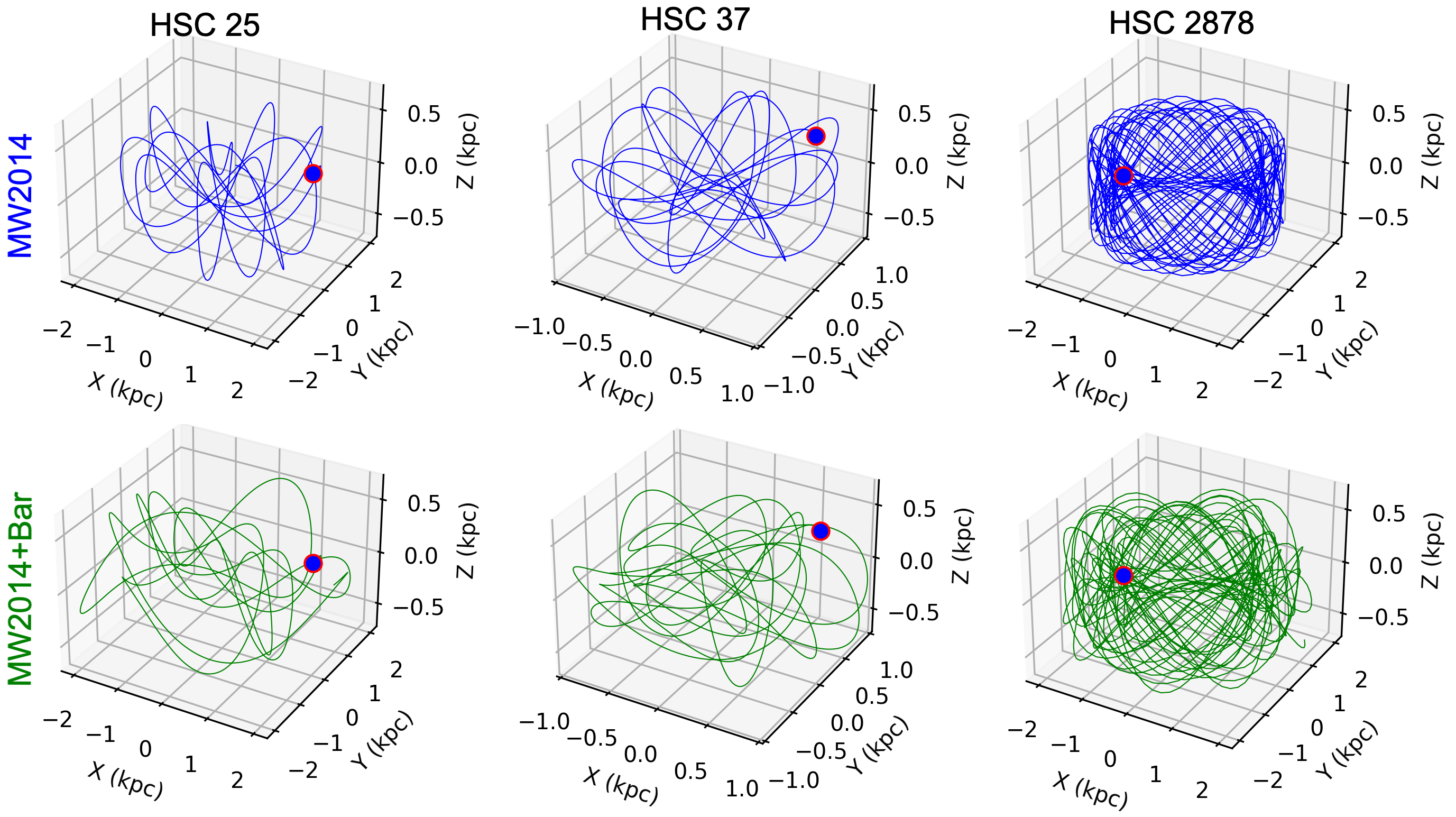} 
\caption{Three-dimensional Galactic orbits of HSC\,25, HSC\,37, and HSC\,2878 shown in the Cartesian $(X, Y, Z)$ coordinate system, centred on the Galactic centre. Blue trajectories correspond to integrations using the axisymmetric \texttt{MWPotential2014} while green trajectories include the non-axisymmetric \texttt{DehnenBarPotential}. The differences between the two models highlight the bar-induced perturbations in the orbital paths, particularly in the inner Galaxy.}
\label{Fig:XYZ} 
\end{figure*}

{In our simulations, we adopted literature-supported parameters with $\Omega_{\rm bar} = 40\,\mathrm{km\,s^{-1}\,kpc^{-1}}$, $R_{\rm bar} = 3.5$\,kpc, a dimensionless amplitude $\alpha = 0.01$, and a bar orientation angle of $25^\circ$ relative to the Sun–Galactic centre line \citep{bland2016, Portail2017}. This composite potential setup allows us to examine the orbital response of clusters located in the dynamically complex inner Galaxy, where the bar can cause significant deviations from circular orbits via resonant trapping, increased eccentricity, or radial migration.

We integrated the orbits of each cluster using both the standard {\sc MWPotential2014} model and the extended version that includes the rotating bar potential. The resulting trajectories are displayed in Figures~\ref{Fig:Orbits} and \ref{Fig:XYZ}, where differences between the two models become especially pronounced near perigalactic passages.

The orbital calculations were performed by adopting key Galactic parameters: the Sun’s circular velocity ($V_{\rm rot} = 220$ km s$^{-1}$), and its vertical offset from the Galactic plane ($Z_{0} = 25 \pm 5$ pc) \citep{Juric2008}. These values were chosen to ensure consistency with the structural and kinematical framework described in the Methods section. For each OC, we considered essential astrometric inputs equatorial coordinates ($\alpha$, $\delta$), heliocentric distances ($d$), proper-motion components ($\mu_{\alpha}\cos\delta$, $\mu_{\delta}$), and radial velocities ($V_{\rm r}$). Based on these parameters, present-day Galactic positions were established, and backwards orbit integrations were carried out using a time resolution of 1 Myr, consistent with the OCs' estimated ages \citep{Sahin2020, Cinar2025, Canbay_2025}. 

The orbital simulations, performed under both the axisymmetric {\sc MWPotential2014} model and its non-axisymmetric extension including the {\tt DehnenBarPotential}, are summarized in Table~\ref{tab:full_parameters} and visualized in Figure~\ref{Fig:Orbits}. For each potential, we report the apogalactic ($R_a$) and perigalactic ($R_p$) distances, representing the farthest and closest points of the cluster's orbit relative to the Galactic centre. The parameter $Z_{\max}$ denotes the maximum vertical displacement from the Galactic plane, while the orbital eccentricity, defined as $e = (R_a - R_p)/(R_a + R_p)$, characterises the orbital shape. Additionally, the azimuthal period $T_{\rm p}$ indicates the time required for a complete orbital revolution around the Galactic centre, serving as a measure of the angular orbital timescale. By comparing results obtained under both potentials, the influence of the Galactic bar on the clusters' orbits is explicitly revealed.

To assess the impact of observational uncertainties on the derived orbital solutions, we performed a Monte Carlo analysis by varying the input velocity components of each OC within their measured uncertainties. For each cluster, multiple orbital integrations were carried out for both the axisymmetric {\sc MWPotential2014} model and the non-axisymmetric configuration, including the {\tt DehnenBarPotential}. The resulting dispersions in the derived orbital parameters (such as $R_{\rm a}$, $R_{\rm p}$, $Z_{\rm max}$, eccentricity, and orbital period) were used to quantify the associated uncertainties. These uncertainty estimates are reported in Table~\ref{tab:full_parameters}, while the corresponding orbital tracks are illustrated in Figures~\ref{Fig:Orbits} and~\ref{Fig:XYZ}. We find that, although velocity uncertainties slightly broaden the parameter ranges, the overall orbital classifications and qualitative dynamical behaviour of the clusters remain unchanged.

\begin{table*}[h]
\centering
\footnotesize
\caption{The evolving, kinematical, and dynamical parameters for the HSC 25, HSC 37, and HSC 2878 OCs.}
\renewcommand{\arraystretch}{1.1}
\begin{tabular}{ll|ccc}
\toprule
&Parameter & HSC 25 & HSC 37 & HSC 2878 \\
\midrule
\multicolumn{5}{c}{Dynamical Parameters} \\
\midrule
&$R_h$ (pc) & 15.74 $\pm$ 3.39 & 8.50 $\pm$ 2.51 & 14.52 $\pm$ 3.28 \\
&$T_{\rm relax}$ (Myr) & 156 $\pm$ 54 & 31 $\pm$ 15 & 214 $\pm$ 76 \\
&$M_{\text{C}}$ ($M_{\odot}$) & 135 & 755  & 204  \\
&$\langle M_C \rangle$ ($M_{\odot}$) & 3.56 &  15.32  & 2.17   \\
\midrule
\multicolumn{5}{c}{Kinematical Parameters} \\
\midrule
&$V_{\rm R}$ km s$^{-1}$) &  -101.09 $\pm$ 35.72 & 41.04 $\pm$ 29.08 & -124.64 $\pm$ 19.43 \\
&$\overline{V_x}$ (km s$^{-1}$) & -193.80 $\pm$ 13.92 & -96.78 $\pm$ 9.84 & -60.23 $\pm$ 7.76 \\
&$\overline{V_y}$ (km s$^{-1}$) & 202.40 $\pm$ 14.23 & 30.23 $\pm$ 5.50 & 209.30 $\pm$ 14.47 \\
&$\overline{V_z}$ (km s$^{-1}$) & -153.54 $\pm$ 12.40 & -145.03 $\pm$ 12.04 & -43.15 $\pm$ 6.57 \\
&$\overline{U}$ (km s$^{-1}$) & -91.81 $\pm$ 9.58 & 49.19 $\pm$ 7.01 & -158.44 $\pm$ 12.59 \\
&$\overline{V}$ (km s$^{-1}$) & -299.92 $\pm$ 17.32 & -169.41 $\pm$ 13.02 & -155.33 $\pm$ 12.46 \\
&$\overline{W}$ (km s$^{-1}$) & 60.91 $\pm$ 7.81 & 13.88 $\pm$ 3.73 & -7.83 $\pm$ 2.80 \\
&$U_{\rm LSR}$ (km s$^{-1}$) & -73.85 $\pm$9.58 & 67.90$\pm$ 7.01& -185.64 $\pm$ 12.59 \\
&$V_{\rm LSR}$ (km s$^{-1}$) & -260.69 $\pm$ 17.32 & -134.04 $\pm$ 13.02 & -128.52$\pm$ 12.47 \\
&$W_{\rm LSR}$ (km s$^{-1}$) & 67.48 $\pm$ 7.81 & 20.45$\pm$ 3.74 & -1.26 $\pm$ 2.81 \\
&$S_{\rm LSR}$ (km s$^{-1}$) & 279.23 $\pm$ 21.28 & 151.64 $\pm$ 15.26 & 225.79 $\pm$ 17.94 \\
\midrule
\multicolumn{5}{c}{Galactic Orbit Parameters} \\
\midrule
&$T_{\rm p}$ (Myr) & 54 $\pm$ 8 & 37 $\pm$ 1 & 59 $\pm$ 1 \\
\multirow{5}{*}{\rotatebox{90}{\underline{MW2014}}}&$Z_{\rm max}$ (kpc) & 0.507 $\pm$ 0.156 & 0.547 $\pm$ 0.010 & 0.396 $\pm$ 0.015 \\
&$R_{\rm a}$ (kpc) & 2.100 $\pm$ 0.459 & 1.312 $\pm$ 0.196 & 1.939 $\pm$ 0.149 \\
&$R_{\rm p}$ (kpc) & 0.199 $\pm$ 0.001& 0.310 $\pm$ 0.131 & 1.445 $\pm$ 0.271 \\
&$R_{\rm m}$ (kpc) & 1.149 $\pm$ 0.230 & 0.811 $\pm$ 0.163 & 1.692 $\pm$ 0.210 \\
&$e$ & 0.827 $\pm$ 0.041 & 0.618 $\pm$ 0.189 & 0.146 $\pm$ 0.131 \\
\hline
\multirow{5}{*}{\rotatebox{90}{\underline{MW2014+Bar}}} & $Z_{\rm max}$ (kpc)  & 0.510 $\pm$ 0.066 & 0.523 $\pm$ 0.021 & 0.436 $\pm$ 0.073 \\
& $R_{\rm a}$ (kpc) & 2.843 $\pm$ 0.557 & 1.407 $\pm$ 0.021 & 2.446 $\pm$ 0.140 \\
& $R_{\rm p}$ (kpc) & 0.220 $\pm$ 0.033 & 0.165 $\pm$ 0.155 & 1.173 $\pm$ 0.344 \\
& $R_{\rm m}$ (kpc) & 1.531 $\pm$ 0.295 & 0.786 $\pm$ 0.088 & 1.810 $\pm$ 0.242 \\
& $e$ & 0.857 $\pm$ 0.050 & 0.790 $\pm$ 0.192 & 0.352 $\pm$ 0.116 \\
\bottomrule
\end{tabular}
\label{tab:full_parameters}
\end{table*}

\section{Summary and Conclusion}
\label{sec4}

Open clusters located near the Galactic centre remain among the least explored stellar systems due to severe extinction, high stellar densities, and strong dynamical perturbations. In this study, we present a comprehensive analysis of three OCs, HSC 25, HSC 37, and HSC 2878, which rank among the closest known OCs to the Galactic centre and have so far lacked detailed characterisation in the literature.

These systems span a wide range of evolutionary stages, from the very young HSC 37 ($\log (t/{\rm yr}) = 7.04$), through the intermediate-age HSC 25 ($\log (t/{\rm yr}) = 8.38$), to the older HSC 2878 ($\log (t/{\rm yr}) = 9.04$), providing a rare opportunity to investigate the age-dependent properties of OCs in the innermost disc. This chronological diversity enables a more comprehensive examination of cluster evolution under the extreme environmental conditions near the Galactic centre. Considering the metallicities of the clusters, their chemical enrichment supports complex formation histories within the Galactic bulge. Binary merger events and other dynamical interactions can enhance metal abundances while also influencing cluster properties such as age dispersion and stellar population characteristics \citep[e.g.,][]{Molero2024}

To refine these OCs' fundamental parameters and membership statistics, we utilise recent data from \citep{Hunt2024} in conjunction with our photometric analysis. While \citet{Hunt2024} reports 18, 55, and 776 members for HSC 25, HSC 37, and HSC 2878, respectively, our revised assessment identifies 44, 55, and 112 members. These discrepancies, particularly striking in the case of HSC 2878, likely stem from differences in membership selection criteria and the handling of background contamination.

The trigonometric parallaxes ($\varpi$) and corresponding parallax-based distances ($d_{\varpi}$) place the OCs at heliocentric distances between approximately 9.3 and 12.8 kpc. However, our isochrone fitting yields systematically smaller and more plausible distances: 7.36 $\pm$ 0.37 kpc for HSC 25, 6.79 $\pm$ 0.18 kpc for HSC 37, and 6.17 $\pm$ 0.22 kpc for HSC 2878. These values are consistent with the geometric distances ($d_{\rm BJ}$) derived by \citet{BailerJones2021} and the $dist50$ estimates from \citet{Hunt2024}. The large discrepancy between $\varpi$-based and isochrone-based distances likely stems from underestimated parallaxes in highly extincted and crowded fields toward the Galactic centre, where Gaia's astrometric precision is significantly compromised.

The amount of extinction and reddening varies significantly among the three clusters. HSC 25 is the most heavily reddened, with a visual extinction of $A_{\rm V} =4.43$ mag and a colour excess of $E(B-V) = 1.41$ mag. HSC 37 exhibits a moderate extinction level ($A_{\rm V} = 2.93$ mag, $E(B-V) = 0.94$ mag), while HSC 2878 is affected by the least amount of dust, with $A_{\rm V} = 3.41$ mag and $E(B-V) = 1.09$ mag. These extinction patterns are consistent with the expected distribution of interstellar dust along lines of sight toward the inner Galaxy. Although substantial differential reddening is evident within each field, we corrected the extinction for each star individually based on its line-of-sight distance, minimising its impact on the distance and age determination.

Cluster mass estimates provide another layer of characterisation. In our analysis, HSC 25 and HSC 2878 display comparable total masses of 135 and 204 $M_{\odot}$, respectively, despite the larger number of members in HSC 2878. This indicates that HSC 2878 hosts a higher fraction of lower-mass stars, yielding a lower average stellar mass ($\langle M_C \rangle = 2.17\ M_{\odot}$) compared to HSC 25 ($\langle M_C \rangle = 3.56\ M_{\odot}$). In contrast, the markedly younger HSC 37 stands out with both the highest total mass (755 $M_{\odot}$) and the highest mean stellar mass ($15.32\ M_{\odot}$), consistent with the presence of massive, luminous stars characteristic of very young stellar populations.

The evolutionary state of each OC can also be assessed through its dynamical relaxation times. For HSC 25, the estimated relaxation time is $156 \pm 54$ Myr, compared to its age of 104 Myr, suggesting that it is still approaching full dynamical relaxation. HSC 2878, with a relaxation time of $214 \pm 76$ Myr and an age of about 1 Gyr, has likely experienced multiple relaxation cycles and can be considered dynamically evolved. In contrast, HSC 37, with a relaxation time of $31 \pm 15$ Myr but an age of only 4 Myr, remains in the very early stages of its dynamical evolution, likely retaining signatures of its initial conditions and being less influenced by long-term internal dynamical processes.

The LSR-corrected velocity components of HSC\,25, HSC\,37, and HSC\,2878 indicate that, despite their common association with the Galactic bulge, they occupy different regions of velocity space, reflecting possible differences in their dynamical histories within the inner Galaxy. HSC\,25 exhibits a high total velocity ($S_{\rm LSR} = 279.23 \pm 21.28$ km~s$^{-1}$) dominated by a large negative $V_{\rm LSR}$ and a substantial vertical component ($W_{\rm LSR} = 67.48 \pm 7.81$ km~s$^{-1}$), suggesting that it follows an orbit with significant excursions above and below the Galactic plane, possibly influenced by past dynamical heating or bar–bulge interactions. HSC\,37 has the lowest total velocity of the sample ($S_{\rm LSR} = 151.64 \pm 15.26$ km~s$^{-1}$), moderate radial motion, and a relatively small vertical velocity, indicating a more planar, less dynamically heated orbit that keeps it closely bound to the bulge region. HSC\,2878, in contrast, is characterised by a large negative $U_{\rm LSR}$ ($-185.64 \pm 12.59$ km~s$^{-1}$), moderate tangential velocity, and negligible vertical motion, pointing to a highly eccentric, low-inclination orbit confined mainly to the bulge. These results show that bulge clusters can display a wide range of orbital shapes and inclinations, shaped by the complex dynamical environment of the inner Milky Way.

 The kinematic properties of the open clusters (OCs), particularly their $S_{\rm LSR}$ values, $279.23 \pm 21.28$, $151.64 \pm 15.26$, and $225.79 \pm 17.94$ km s$^{-1}$ for HSC 25, HSC 37, and HSC 2878, respectively, are considerably higher than the velocity dispersions typically observed among thin-disc stars. These high space motions suggest that the clusters may not be members of the canonical thin disc population, warranting further analysis of their orbits within both axisymmetric and barred Galactic potentials. Orbital integrations in the axisymmetric MW2014 model reveal that all three OCs remain relatively confined to the Galactic plane, with maximum vertical distances $Z_{\rm max}$ ranging from $0.396 \pm 0.015$~kpc (HSC 2878) to $0.547 \pm 0.010$~kpc (HSC 37). These vertical distances remain moderate even in the barred potential (MW2014+Bar), where $Z_{\rm max}$ values vary between $0.436 \pm 0.073$~kpc and $0.523 \pm 0.021$~kpc.

 Including the barred Galactic potential in the MW2014 potential model leads to noticeable variations in the derived orbital parameters for the analyzed clusters. In particular, the apocentric distance ($R_{\rm a}$) shows the most pronounced relative increase, reaching $\sim35\%$ for HSC 25, $\sim7\%$ for HSC 37, and $\sim26\%$ for HSC 2878 compared to the axisymmetric case. Moderate changes are also observed in the mean orbital radius ($R_{\rm m}$), with increases of $\sim33\%$, $-3\%$, and $\sim7\%$ for the respective clusters. The eccentricities ($e$) exhibit enhancements of $\sim4\%$, $\sim28\%$, and more than twofold for HSC 2878, indicating that the bar component can significantly alter orbital shapes, particularly for low-eccentricity orbits. The vertical excursion ($Z_{\rm max}$) remains relatively stable, with variations within a few percent, suggesting that the bar’s influence is predominantly confined to the radial orbital structure rather than the vertical dynamics.

However, their radial orbital properties are notably distinct. HSC 25 exhibits a highly eccentric orbit, with $e = 0.827 \pm 0.041$ (MW2014) and $e = 0.857 \pm 0.050$ (MW2014+Bar), and plunges from a small pericentric distance ($R_{\rm p} = 0.199 \pm 0.001$~kpc) to a relatively extended apocentre ($R_{\rm a} = 2.843 \pm 0.557$~kpc). HSC 37 follows a similarly elliptical path with $e = 0.618 \pm 0.189$ in MW2014, increasing to $e = 0.790 \pm 0.192$ under the barred model. In contrast, HSC 2878 remains more tightly bound to the inner disc, showing the lowest eccentricities among the three ($e = 0.146 \pm 0.131$ in MW2014 and $e = 0.352 \pm 0.116$ in MW2014+Bar), and a narrower radial range. In terms of orbital time-scales, the clusters complete one revolution around the Galactic centre within approximately $54 \pm 8$ Myr for HSC 25, $37 \pm 1$ Myr for HSC 37, and $59 \pm 1$ Myr for HSC 2878, consistent with their varying orbital sizes and shapes.

Taken together, these results suggest that while all three clusters are moderately confined vertically, their planar orbital properties, particularly the high eccentricities and inner-Galaxy plunging orbits of HSC 25 and HSC 37, imply that they may have originated from, or interacted with, dynamically hotter Galactic components, or even formed in non-standard disc environments.

HSC 2878, on the other hand, shows signs of dynamical quiescence. Its low eccentricity and confinement to the Galactic plane, despite its advanced age, suggest it has avoided the kind of gravitational perturbations, such as interactions with molecular clouds or spiral arms, that typically increase orbital eccentricity over time \citep[e.g.,][]{Wielen1977, Quillen2001, AumerBinney2009}. The orbital and kinematic properties of HSC\,25, HSC\,37, and HSC\,2878 indicate that, although all three are currently confined to the Galactic bulge, they have followed distinct dynamical paths. HSC\,25 and HSC\,2878 show signatures of dynamically heated or possibly accreted origins, whereas HSC\,37 exhibits a cooler, more planar motion consistent with in-situ bulge formation.

These findings support the view that the three OCs do not share a common dynamical past and instead may represent a diverse population shaped by distinct formation environments and evolutionary processes. Future high-resolution spectroscopic studies of their chemical abundances, coupled with orbital integrations in time-dependent, non-axisymmetric Galactic potentials, including bar and spiral arm perturbations, will be essential to constrain their origins and dynamical classifications further.

\section{Acknowledgments}

We would like to thank the anonymous referee for their helpful feedback and recommendations, which have contributed meaningfully to the improvement of our manuscript. This study presents results derived from the European Space Agency (ESA) space mission Gaia. The data from $Gaia$ are processed by the $Gaia$ Data Processing and Analysis Consortium (DPAC). Financial support for DPAC is provided by national institutions, primarily those participating in the $Gaia$ Multi-Lateral Agreement (MLA). For additional information, the official $Gaia$ mission website can be accessed at \url{https://www.cosmos.esa.int/gaia}, and the $Gaia$ archive is available at \url{https://archives.esac.esa.int/gaia}. The authors would like to express their gratitude to the Deanship of Scientific Research at Northern Border University, Arar, KSA, for funding this research under project number "NBU-FFR-2026-237-04".

\bibliographystyle{elsarticle-harv} 


\appendix
\input{Appendix_Members}

\end{document}

%% file: Appendix_Members.tex
\onecolumn
\section*{Appendix A. List of Cluster Members}

\begin{scriptsize} 
\setlength{\tabcolsep}{4pt} 
\begin{longtable}{ccccccc}
\caption{Most probable members identified for HSC 25, HSC 37, and HSC 2878.} \label{tab:members_list} \\
\hline
Order & Source ID & $\alpha$ & $\delta$ & $G$ & $R_{\rm dist}$ & $P$ \\
&   & (deg) & (deg) & (mag) & &  \\
\hline
\multicolumn{7}{c}{HSC 25} \\
\hline
1  & 4061893199113258752 & 265.4063 & -25.7219 & 16.58 & 1.46  & 0.57 \\
2  & 4061892511907882752 & 265.3709 & -25.7297 & 16.18 & 2.02  & 0.53 \\
3  & 4061894745291288064 & 265.4095 & -25.6587 & 17.50 & 2.75  & 0.86 \\
4  & 4061893714509852672 & 265.4497 & -25.6971 & 17.62 & 3.15  & 0.63 \\
5  & 4061941307090339584 & 265.3388 & -25.6648 & 18.27 & 3.61  & 0.73 \\
6  & 4061891554171656704 & 265.3665 & -25.7647 & 19.24 & 4.02  & 0.51 \\
7  & 4061892550662103680 & 265.3209 & -25.7396 & 17.27 & 4.45  & 0.81 \\
8  & 4067899526284562432 & 265.4623 & -25.6431 & 18.70 & 5.19  & 0.71 \\
9  & 4061941959956466176 & 265.3959 & -25.6124 & 17.67 & 5.36  & 0.69 \\
10 & 4061891962151801600 & 265.3212 & -25.7718 & 17.90 & 5.68  & 0.63 \\
11 & 4061891210573332736 & 265.3955 & -25.7991 & 16.29 & 5.86  & 0.65 \\
12 & 4061889934927280384 & 265.4769 & -25.7645 & 15.23 & 5.96  & 0.65 \\
13 & 4061942234803932160 & 265.3765 & -25.5872 & 16.97 & 6.91  & 0.88 \\
14 & 4061892099601587712 & 265.2803 & -25.7648 & 18.18 & 7.11  & 0.72 \\
15 & 4067947011429748864 & 265.3986 & -25.5822 & 17.77 & 7.17  & 0.80 \\
16 & 4061890278524824320 & 265.5349 & -25.7388 & 16.33 & 8.06  & 0.57 \\
17 & 4061946186175148032 & 265.3126 & -25.5482 & 15.19 & 10.15 & 0.77 \\
18 & 4067947385006693504 & 265.4735 & -25.5340 & 16.60 & 10.99 & 0.54 \\
19 & 4067948385819404928 & 265.4307 & -25.5216 & 17.31 & 11.01 & 0.93 \\
20 & 4061886675146119936 & 265.4450 & -25.8816 & 16.22 & 11.18 & 0.80 \\
21 & 4061943437425557888 & 265.1791 & -25.6830 & 17.76 & 11.54 & 0.99 \\
22 & 4067895815339514112 & 265.6164 & -25.6471 & 16.88 & 12.59 & 0.97 \\
23 & 4061884476059729536 & 265.2639 & -25.8806 & 18.70 & 12.77 & 0.53 \\
24 & 4061884269901201408 & 265.2710 & -25.8873 & 19.05 & 12.91 & 0.85 \\
25 & 4067900934943307904 & 265.5895 & -25.5632 & 18.35 & 13.55 & 0.98 \\
26 & 4061880352953047936 & 265.4239 & -25.9399 & 16.04 & 14.41 & 0.87 \\
27 & 4061932717185725696 & 265.1553 & -25.8139 & 19.27 & 14.43 & 0.75 \\
28 & 4061880765208864768 & 265.3067 & -25.9670 & 18.29 & 16.57 & 0.93 \\
29 & 4067896262013746944 & 265.7015 & -25.6659 & 18.31 & 16.89 & 0.75 \\
30 & 4067881040741568896 & 265.6795 & -25.8308 & 18.02 & 17.38 & 0.71 \\
31 & 4061938523962039296 & 265.0704 & -25.6703 & 18.68 & 17.47 & 0.79 \\
32 & 4061932373588471936 & 265.0757 & -25.7829 & 16.62 & 17.75 & 0.58 \\
33 & 4061956837726296064 & 265.1023 & -25.5602 & 16.98 & 17.80 & 0.93 \\
34 & 4067949961988029568 & 265.5405 & -25.4280 & 16.58 & 18.29 & 0.50 \\
35 & 4061959620835955584 & 265.1901 & -25.4511 & 19.33 & 18.57 & 0.77 \\
36 & 4067897017928563584 & 265.7271 & -25.6227 & 18.10 & 18.75 & 0.64 \\
37 & 4061873137339617920 & 265.5121 & -25.9988 & 18.78 & 18.99 & 0.67 \\
38 & 4061957108227340416 & 265.0806 & -25.5389 & 18.51 & 19.45 & 0.68 \\
39 & 4061883131702799104 & 265.1403 & -25.9446 & 16.89 & 19.92 & 0.54 \\
40 & 4061890278524824320 & 265.5349 & -25.7388 & 16.33 & 19.97 & 0.94 \\
41 & 4061892511907882752 & 265.3709 & -25.7297 & 16.18 & 19.97 & 0.87 \\
42 & 4061892550662103680 & 265.3209 & -25.7396 & 17.27 & 19.97 & 0.82 \\
43 & 4061941959956466176 & 265.3959 & -25.6124 & 17.67 & 19.97 & 0.82 \\
44 & 4061942234803932160 & 265.3765 & -25.5873 & 16.97 & 19.98 & 0.85\\
\hline 
\hline
\multicolumn{7}{c}{HSC 37} \\
\hline
1  & 4116297863470964864 & 264.3393 & -24.3909 & 17.72 & 1.24  & 0.53 \\
2  & 4116296725211705856 & 264.3593 & -24.4365 & 16.02 & 1.86  & 0.63 \\
3  & 4116297485513746304 & 264.3997 & -24.3811 & 14.66 & 2.85  & 0.62 \\
4  & 4116296523441351040 & 264.2847 & -24.4011 & 16.51 & 3.85  & 0.75 \\
5  & 4116292155366332928 & 264.3874 & -24.4797 & 17.93 & 4.78  & 0.68 \\
6  & 4116292125394285312 & 264.3724 & -24.4983 & 15.53 & 5.64  & 0.56 \\
7  & 4116297382340345472 & 264.4557 & -24.3732 & 18.00 & 5.84  & 0.67 \\
8  & 4116308201364443136 & 264.2470 & -24.4073 & 17.56 & 5.90  & 0.59 \\
9  & 4116293946460332928 & 264.4693 & -24.4436 & 16.00 & 6.64  & 0.76 \\
10 & 4116291575543116416 & 264.3296 & -24.5359 & 18.71 & 7.94  & 0.97 \\
11 & 4116292258445416576 & 264.4678 & -24.5275 & 15.60 & 9.56  & 0.59 \\
12 & 4116309373893519488 & 264.1895 & -24.3517 & 19.82 & 9.61  & 0.88 \\
13 & 4110291093956974080 & 264.1933 & -24.4706 & 19.33 & 9.66  & 0.54 \\
14 & 4116299336552012416 & 264.5403 & -24.3934 & 17.34 & 10.15 & 0.73 \\
15 & 4116299375202774656 & 264.5542 & -24.3835 & 17.70 & 10.96 & 0.80 \\
16 & 4116300612154452480 & 264.5483 & -24.3477 & 17.53 & 11.12 & 0.99 \\
17 & 4116291678717750016 & 264.3315 & -24.5123 & 16.40 & 11.12 & 0.89 \\
18 & 4116317719016015104 & 264.3607 & -24.2156 & 18.70 & 11.41 & 0.93 \\
19 & 4116292395884419840 & 264.4837 & -24.5094 & 14.44 & 11.41 & 0.52 \\
20 & 4116296415974109568 & 264.2700 & -24.4237 & 16.93 & 11.95 & 0.63 \\
21 & 4116296415974136960 & 264.2746 & -24.4177 & 17.15 & 12.16 & 0.83 \\
22 & 4116296690868885760 & 264.3881 & -24.4368 & 17.78 & 12.31 & 0.84 \\
23 & 4116296793931220608 & 264.3913 & -24.4265 & 16.47 & 12.39 & 0.82 \\
24 & 4116297343687195008 & 264.4347 & -24.3902 & 18.72 & 12.81 & 0.76 \\
25 & 4068251954064951168 & 264.4666 & -24.5936 & 18.30 & 12.81 & 0.59 \\
26 & 4116302501941006080 & 264.5595 & -24.3010 & 18.87 & 12.82 & 0.96 \\
27 & 4116302261519760512 & 264.5974 & -24.3077 & 18.09 & 13.24 & 0.94 \\
28 & 4068254187448027648 & 264.5726 & -24.5030 & 18.51 & 13.24 & 0.53 \\
29 & 4116302982976940800 & 264.6237 & -24.2634 & 18.73 & 13.39 & 0.85 \\
30 & 4116315210755065216 & 264.2465 & -24.2055 & 18.26 & 13.39 & 0.52 \\
31 & 4116297034449477632 & 264.3533 & -24.4072 & 15.14 & 13.73 & 0.65 \\
32 & 4068251473028317056 & 264.4500 & -24.6184 & 18.80 & 13.78 & 0.78 \\
33 & 4116305491238710656 & 264.5679 & -24.2751 & 18.59 & 14.03 & 0.87 \\
34 & 4068251438668414336 & 264.4451 & -24.6283 & 18.28 & 14.23 & 0.71 \\
35 & 4110285042307239296 & 264.2946 & -24.6367 & 18.73 & 14.25 & 0.99 \\
36 & 4116305074629133952 & 264.4995 & -24.2516 & 17.81 & 14.25 & 0.97 \\
37 & 4116318925900274304 & 264.4420 & -24.1788 & 18.20 & 14.42 & 0.67 \\
38 & 4116305933622762112 & 264.5406 & -24.2268 & 17.73 & 14.77 & 0.88 \\
39 & 4116305628677630720 & 264.5834 & -24.2634 & 18.19 & 15.13 & 0.57 \\
40 & 4116319922429152512 & 264.4501 & -24.1404 & 18.06 & 15.40 & 0.74 \\
41 & 4110285012282893696 & 264.2741 & -24.6521 & 16.10 & 15.43 & 0.78 \\
42 & 4116320678343548544 & 264.4470 & -24.0812 & 17.80 & 15.43 & 0.62 \\
43 & 4116303051832779008 & 264.5981 & -24.2558 & 17.52 & 16.05 & 0.96 \\
44 & 4116313359717558656 & 264.0819 & -24.2898 & 20.32 & 16.47 & 0.71 \\
45 & 4116302330140804096 & 264.6336 & -24.2971 & 19.21 & 16.56 & 0.86 \\
46 & 4110284668648646272 & 264.3093 & -24.6789 & 18.16 & 16.58 & 0.86 \\
47 & 4116321807825166464 & 264.3377 & -24.1210 & 18.43 & 17.11 & 0.67 \\
48 & 4116307484200635136 & 264.5251 & -24.1652 & 17.42 & 17.17 & 0.97 \\
49 & 4116307587185626752 & 264.5395 & -24.1630 & 17.86 & 17.72 & 0.96 \\
50 & 4068249411508299008 & 264.5781 & -24.6204 & 17.59 & 17.73 & 0.82 \\
51 & 4116301677304415488 & 264.6741 & -24.3132 & 18.41 & 18.30 & 0.92 \\
52 & 4068259577590439936 & 264.6853 & -24.4830 & 17.78 & 18.63 & 0.78 \\
53 & 4110308239466721280 & 264.0139 & -24.3644 & 17.72 & 18.80 & 0.59 \\
54 & 4068249342720707840 & 264.6126 & -24.6175 & 17.65 & 18.96 & 0.73 \\
55 & 4116320403369755648 & 264.4661 & -24.1064 & 19.28 & 18.96 & 0.85\\
\hline 
\hline
\multicolumn{7}{c}{HSC 2878} \\
\hline
1   & 5976514135629669760 & 256.1042 & -37.0103 & 17.42 & 0.11  & 0.62 \\
2   & 5976701671108223360 & 256.0770 & -37.0171 & 16.65 & 1.25  & 0.50 \\
3   & 5976513620263457280 & 256.0645 & -37.0400 & 18.87 & 2.51  & 0.52 \\
4   & 5976519805019042304 & 256.1616 & -36.9960 & 17.90 & 2.99  & 0.50 \\
5   & 5976707817166021760 & 256.1196 & -36.9516 & 15.62 & 3.65  & 0.50 \\
6   & 5976513001799309696 & 256.1650 & -37.0485 & 17.91 & 3.78  & 0.78 \\
7   & 5976702186504436224 & 256.0727 & -36.9493 & 16.64 & 3.94  & 0.58 \\
8   & 5976702908059382144 & 256.0535 & -36.9491 & 18.33 & 4.36  & 0.76 \\
9   & 5976702392618718848 & 256.0127 & -36.9826 & 17.52 & 4.60  & 0.69 \\
10  & 5976702942374695296 & 256.0625 & -36.9390 & 18.78 & 4.70  & 0.58 \\
11  & 5976512795601081216 & 256.1752 & -37.0660 & 19.84 & 4.83  & 0.54 \\
12  & 5976699712554813056 & 256.0037 & -37.0663 & 18.12 & 5.77  & 0.50 \\
13  & 5976518769884329600 & 256.2317 & -37.0348 & 17.50 & 6.38  & 0.63 \\
14  & 5976509771942057088 & 256.2019 & -37.0840 & 18.27 & 6.50  & 0.54 \\
15  & 5976699884353567232 & 255.9820 & -37.0645 & 19.87 & 6.59  & 0.50 \\
16  & 5976519457078122624 & 256.2427 & -36.9865 & 17.82 & 6.90  & 0.68 \\
17  & 5976519152183954432 & 256.2469 & -37.0123 & 17.62 & 6.94  & 0.61 \\
18  & 5976702491405715456 & 255.9596 & -36.9821 & 16.26 & 7.03  & 0.63 \\
19  & 5976519250920735360 & 256.2491 & -36.9886 & 16.94 & 7.18  & 0.54 \\
20  & 5976519186543228288 & 256.2619 & -37.0034 & 16.29 & 7.68  & 0.63 \\
21  & 5976512138454675840 & 256.1376 & -37.1413 & 18.00 & 8.02  & 0.71 \\
22  & 5976709436365594752 & 256.1020 & -36.8742 & 18.47 & 8.19  & 0.50 \\
23  & 5976512035377017728 & 256.1283 & -37.1475 & 18.28 & 8.30  & 0.56 \\
24  & 5976512001017260288 & 256.1100 & -37.1559 & 18.38 & 8.72  & 0.50 \\
25  & 5976509084777223936 & 256.1298 & -37.1548 & 17.61 & 8.75  & 0.50 \\
26  & 5976509084777212288 & 256.1381 & -37.1544 & 19.40 & 8.79  & 0.69 \\
27  & 5976709264613265408 & 256.0815 & -36.8650 & 16.57 & 8.80  & 0.56 \\
28  & 5976521145048449536 & 256.2559 & -36.9297 & 17.41 & 8.84  & 0.70 \\
29  & 5976510729705387904 & 256.0441 & -37.1540 & 17.45 & 9.03  & 0.61 \\
30  & 5976509325265076480 & 256.1905 & -37.1438 & 20.35 & 9.04  & 0.50 \\
31  & 5976705622477958400 & 255.9332 & -36.9420 & 18.67 & 9.08  & 0.67 \\
32  & 5976515303861745792 & 256.2437 & -37.1118 & 20.36 & 9.10  & 0.54 \\
33  & 5976706309629640576 & 256.0000 & -36.8782 & 17.81 & 9.34  & 0.57 \\
34  & 5976521935322475904 & 256.2499 & -36.9085 & 19.82 & 9.38  & 0.50 \\
35  & 5976509119101964032 & 256.1559 & -37.1636 & 19.13 & 9.53  & 0.50 \\
36  & 5976709638231918464 & 256.0851 & -36.8505 & 16.79 & 9.65  & 0.66 \\
37  & 5976511386845424896 & 255.9759 & -37.1430 & 19.13 & 9.97  & 0.50 \\
38  & 5976514719777568000 & 256.2837 & -37.0926 & 18.52 & 9.99  & 0.75 \\
39  & 5976515888008710144 & 256.3058 & -37.0501 & 17.08 & 10.05 & 0.57 \\
40  & 5976511386845440128 & 255.9709 & -37.1426 & 18.65 & 10.10 & 0.64 \\
41  & 5976699403321193216 & 255.9172 & -37.0989 & 20.11 & 10.31 & 0.50 \\
42  & 5976510493526489088 & 256.0907 & -37.1842 & 14.28 & 10.42 & 0.56 \\
43  & 5976511112001942784 & 256.0023 & -37.1679 & 16.21 & 10.57 & 0.56 \\
44  & 5976517812162823552 & 256.3195 & -36.9759 & 19.10 & 10.63 & 0.88 \\
45  & 5976508878588286080 & 256.1870 & -37.1745 & 19.63 & 10.64 & 0.64 \\
46  & 5976704145009189760 & 255.8789 & -36.9926 & 16.69 & 10.74 & 0.50 \\
47  & 5976508122704928256 & 256.1978 & -37.1739 & 17.82 & 10.81 & 0.50 \\
48  & 5976710260996681472 & 256.1723 & -36.8386 & 18.64 & 10.87 & 0.63 \\
49  & 5976508771200255360 & 256.1417 & -37.1904 & 16.82 & 10.94 & 0.53 \\
50  & 5976511112001956480 & 255.9896 & -37.1698 & 14.40 & 10.95 & 0.75 \\
51  & 5976507538554018944 & 256.1027 & -37.1961 & 19.36 & 11.12 & 0.72 \\
52  & 5976712734946894976 & 256.0604 & -36.8275 & 15.79 & 11.17 & 0.54 \\
53  & 5976515162113372672 & 256.3256 & -37.0650 & 19.38 & 11.20 & 0.85 \\
54  & 5976522347608711040 & 256.2621 & -36.8631 & 20.38 & 11.73 & 0.88 \\
55  & 5976517532933798272 & 256.3462 & -36.9963 & 17.70 & 11.73 & 0.63 \\
56  & 5976706825025860224 & 255.9071 & -36.8913 & 19.56 & 11.77 & 0.50 \\
57  & 5976517949592644608 & 256.3404 & -36.9515 & 17.44 & 11.97 & 0.73 \\
58  & 5976700365390156672 & 255.8656 & -37.0775 & 20.04 & 12.00 & 0.89 \\
59  & 5976510218613863424 & 256.0218 & -37.2016 & 19.86 & 12.08 & 0.50 \\
60  & 5976705004002716416 & 255.8600 & -36.9512 & 16.33 & 12.14 & 0.50 \\
61  & 5976605154573384448 & 255.9148 & -37.1478 & 20.17 & 12.16 & 0.50 \\
62  & 5976605360731824256 & 255.9104 & -37.1455 & 20.19 & 12.23 & 0.53 \\
63  & 5976508221444115328 & 256.2312 & -37.1865 & 18.95 & 12.23 & 0.61 \\
64  & 5976522721256048768 & 256.2653 & -36.8518 & 18.44 & 12.34 & 0.52 \\
65  & 5976516884405080576 & 256.3596 & -37.0067 & 19.15 & 12.35 & 0.71 \\
66  & 5976516678282154624 & 256.3600 & -37.0342 & 16.97 & 12.44 & 0.52 \\
67  & 5976510111249265664 & 256.0149 & -37.2062 & 17.97 & 12.45 & 0.62 \\
68  & 5976509978130811392 & 256.0559 & -37.2163 & 20.03 & 12.53 & 0.50 \\
69  & 5976707026891755392 & 255.8992 & -36.8772 & 16.55 & 12.60 & 0.94 \\
70  & 5976711119990305664 & 256.1586 & -36.7982 & 18.30 & 13.04 & 0.91 \\
71  & 5976706928148136832 & 255.8743 & -36.8914 & 18.58 & 13.05 & 0.50 \\
72  & 5976518327514357376 & 256.3641 & -36.9463 & 18.47 & 13.15 & 0.92 \\
73  & 5976605326368710528 & 255.8863 & -37.1554 & 20.46 & 13.48 & 0.65 \\
74  & 5976606146700353536 & 255.8644 & -37.1355 & 18.36 & 13.61 & 0.58 \\
75  & 5976518464953395200 & 256.3700 & -36.9236 & 18.90 & 13.87 & 0.50 \\
76  & 5976506743979176064 & 256.0878 & -37.2492 & 17.70 & 14.33 & 0.54 \\
77  & 5976417103739787648 & 255.9489 & -37.2167 & 18.58 & 14.37 & 0.50 \\
78  & 5976718713491140992 & 255.8523 & -36.8746 & 18.45 & 14.49 & 0.64 \\
79  & 5976704832204061824 & 255.8013 & -36.9644 & 17.29 & 14.67 & 0.50 \\
80  & 5976606077980856576 & 255.8410 & -37.1481 & 18.60 & 14.97 & 0.95 \\
81  & 5976496165484562944 & 256.2922 & -37.2207 & 20.14 & 15.55 & 0.59 \\
82  & 5976502865631133568 & 256.3929 & -37.1275 & 20.01 & 15.60 & 0.88 \\
83  & 5976415969885811072 & 256.0156 & -37.2653 & 17.99 & 15.82 & 0.91 \\
84  & 5976495955019241600 & 256.2555 & -37.2448 & 18.06 & 15.85 & 0.54 \\
85  & 5976506091139537152 & 256.1802 & -37.2678 & 19.02 & 15.87 & 0.62 \\
86  & 5976610823980364160 & 255.7649 & -37.0232 & 20.02 & 16.16 & 0.50 \\
87  & 5976716720626283392 & 255.7674 & -36.9412 & 17.47 & 16.57 & 0.62 \\
88  & 5976610823930633344 & 255.7520 & -37.0163 & 17.86 & 16.77 & 0.93 \\
89  & 5976605940541912064 & 255.7923 & -37.1418 & 18.15 & 16.78 & 0.62 \\
90  & 5976501903560969344 & 256.3923 & -37.1708 & 18.67 & 16.90 & 0.63 \\
91  & 5976622678041516288 & 255.7483 & -36.9906 & 20.40 & 16.99 & 0.94 \\
92  & 5976605601246736896 & 255.8187 & -37.1820 & 18.28 & 17.01 & 0.54 \\
93  & 5976502006637520768 & 256.4324 & -37.1373 & 19.42 & 17.55 & 0.50 \\
94  & 5976716789397507968 & 255.7414 & -36.9482 & 19.60 & 17.68 & 0.55 \\
95  & 5976506503458792704 & 256.0688 & -37.3058 & 17.95 & 17.78 & 0.94 \\
96  & 5976725581144747392 & 255.9149 & -36.7549 & 17.50 & 17.78 & 0.50 \\
97  & 5976416755817554816 & 255.8898 & -37.2564 & 18.45 & 17.89 & 0.84 \\
98  & 5976542134506734848 & 256.4331 & -36.8696 & 17.65 & 18.00 & 0.93 \\
99  & 5976501285114858880 & 256.3659 & -37.2273 & 15.72 & 18.12 & 0.65 \\
100 & 5976541932695419264 & 256.4402 & -36.8752 & 17.02 & 18.14 & 0.65 \\
101 & 5976495650083822848 & 256.2674 & -37.2845 & 18.84 & 18.23 & 0.50 \\
102 & 5976541108031789824 & 256.4652 & -36.9107 & 19.74 & 18.42 & 0.89 \\
103 & 5976734312868607232 & 256.2993 & -36.7472 & 16.39 & 18.43 & 0.92 \\
104 & 5976416450906116096 & 255.8968 & -37.2754 & 20.27 & 18.67 & 0.92 \\
105 & 5976623021637925120 & 255.7171 & -36.9524 & 18.75 & 18.78 & 0.53 \\
106 & 5976725752947556736 & 255.9533 & -36.7198 & 18.37 & 18.86 & 0.84 \\
107 & 5976501044596678144 & 256.3805 & -37.2368 & 17.74 & 19.02 & 0.51 \\
108 & 5976715032711487104 & 256.1908 & -36.7013 & 18.48 & 19.05 & 0.86 \\
109 & 5976542581183416960 & 256.4309 & -36.8329 & 15.18 & 19.05 & 0.87 \\
110 & 5976715032711489408 & 256.1951 & -36.7009 & 19.45 & 19.12 & 0.50 \\
111 & 5976717820142606208 & 255.7316 & -36.8909 & 18.48 & 19.16 & 0.81 \\
112 & 5976726581926205312 & 255.9259 & -36.7103 & 19.56 & 19.91 & 0.91\\
\hline
\end{longtable}
\end{scriptsize}

\twocolumn